\documentclass[twocolumn]{aastex62}

\usepackage{graphicx}
\usepackage[english]{babel}
\usepackage{amssymb}
\usepackage{xspace}
\usepackage{amsmath}
\usepackage{color}

\def\hi{\ifmmode {\mbox H{\scshape i}}\else H{\scshape i}\fi\xspace}
\def\hii{\ifmmode {\mbox H{\scshape ii}}\else H{\scshape ii}\fi\xspace}
\def\h2{\ifmmode {\mbox H$_2$}\else H$_2$\fi\xspace}

\newcommand{\gp }{\color{black}}

\submitjournal{ApJ}

\shorttitle{The theoretical context of the ASPECS}

\shortauthors{G. Popping et al.}

\begin{document}
\title{The ALMA Spectroscopic Survey in the HUDF: the molecular gas content of galaxies \\and tensions with IllustrisTNG and the Santa Cruz SAM}

\correspondingauthor{Gerg\"o Popping}
\email{popping@mpia.de}

\author[0000-0003-1151-4659]{Gerg\"{o} Popping}
\affil{Max Planck Institute f\"ur Astronomie, K\"onigstuhl 17, 69117 Heidelberg, Germany}

\author{Annalisa Pillepich}
\affil{Max Planck Institute f\"ur Astronomie, K\"onigstuhl 17, 69117 Heidelberg, Germany}

\author{Rachel S. Somerville}
\affil{Center for Computational Astrophysics, Flatiron Institute, 162 5th Ave, New York, NY 10010, USA}
\affil{Department of Physics and Astronomy, Rutgers, The State University of New Jersey, 136 Frelinghuysen Rd,
Piscataway, NJ 08854, USA}

\author[0000-0002-2662-8803]{Roberto Decarli}
\affil{INAF—Osservatorio di Astrofisica e Scienza dello Spazio, via Gobetti 93/3, I-40129, Bologna, Italy}

\author[0000-0003-4793-7880]{Fabian Walter}
\affil{Max Planck Institute f\"ur Astronomie, K\"onigstuhl 17, 69117 Heidelberg, Germany}
\affil{National Radio Astronomy Observatory, Pete V. Domenici Array Science Center, P.O. Box O, Socorro, NM 87801, USA}

\author{Manuel Aravena}
\affil{N\'{u}cleo de Astronom\'{\i}a, Facultad de Ingenier\'{\i}a, Universidad Diego Portales, Av. Ej\'{e}rcito 441, Santiago, Chile}

\author{Chris Carilli}
\affil{National Radio Astronomy Observatory, Pete V. Domenici Array Science Center, P.O. Box O, Socorro, NM 87801, USA}
\affil{Battcock Centre for Experimental Astrophysics, Cavendish Laboratory,
Cambridge CB3 0HE, UK}

\author{Pierre Cox}
\affil{Institut d’astrophysique de Paris, Sorbonne Universit\'e, CNRS, UMR 7095, 98 bis bd Arago, 7014 Paris, France}

\author{Dylan Nelson}
\affil{Max-Planck-Institut f\"ur Astrophysik, Karl-Schwarzschild-Str. 1, 85741 Garching, Germany}

\author[0000-0001-9585-1462]{Dominik Riechers}
\affil{Cornell University, 220 Space Sciences Building, Ithaca, NY 14853, USA}
\affil{Max Planck Institute f\"ur Astronomie, K\"onigstuhl 17, 69117 Heidelberg, Germany}

\author{Axel Weiss}
\affil{Max-Planck-Institut f\"ur Radioastronomie, Auf dem H\"ugel 69, 53121 Bonn, Germany}

\author{Leindert Boogaard}
\affil{Leiden Observatory, Leiden University, PO Box 9513, NL-2300 RA Leiden, The Netherlands}

\author{Richard Bouwens}
\affil{Leiden Observatory, Leiden University, PO Box 9513, NL-2300 RA Leiden, The Netherlands}

\author{Thierry Contini}
\affil{Institut de Recherche en Astrophysique et Planetologie (IRAP), Universit\'e de Toulouse, CNRS, UPS, 31400 Toulouse, France}

\author{Paulo C.~Cortes}
\affil{Joint ALMA Observatory - ESO, Av. Alonso de C\'ordova, 3104, Santiago, Chile}
\affil{National Radio Astronomy Observatory, 520 Edgemont Rd, Charlottesville, VA, 22903, USA} 

\author{Elisabete da Cunha}
\affil{Research School of Astronomy and Astrophysics, Australian National University, Canberra, ACT 2611, Australia}

\author{Emanuele Daddi}
\affil{Laboratoire AIM, CEA/DSM-CNRS-Universite Paris Diderot, Irfu/Service d'Astrophysique, CEA Saclay, Orme des Merisiers, 91191 Gif-sur-Yvette cedex, France}

\author{Tanio D\'iaz-Santos}
\affil{N\'{u}cleo de Astronom\'{\i}a, Facultad de Ingenier\'{\i}a, Universidad Diego Portales, Av. Ej\'{e}rcito 441, Santiago, Chile}

\author{Benedikt Diemer}
\affil{Harvard-Smithsonian Center for Astrophysics, 60 Garden Street, Cambridge, MA, 02138, USA}

\author[0000-0003-3926-1411]{Jorge Gonz\'alez-L\'opez}
\affil{N\'ucleo de Astronom\'ia de la Facultad de Ingenier\'ia y Ciencias, Universidad Diego Portales, Av. Ej\'ercito Libertador 441, Santiago, Chile}
\affil{Instituto de Astrof\'{\i}sica, Facultad de F\'{\i}sica, Pontificia Universidad Cat\'olica de Chile Av. Vicu\~na Mackenna 4860, 782-0436 Macul, Santiago, Chile}

\author{Lars Hernquist}
\affil{Harvard-Smithsonian Center for Astrophysics, 60 Garden Street, Cambridge, MA, 02138, USA}

\author{Rob Ivison}
\affil{European Southern Observatory, Karl-Schwarzschild-Strasse 2, 85748, Garching, Germany}
\affil{Institute for Astronomy, University of Edinburgh, Royal Observatory, Blackford Hill, Edinburgh EH9 3HJ}

\author{Olivier Le F\`evre}
\affil{Aix Marseille Universit\'e, CNRS, LAM (Laboratoire d'Astrophysique de Marseille), UMR 7326, F-13388 Marseille, France}

\author{Federico Marinacci}
\affil{Kavli Institute for Astrophysics and Space Research, Department of Physics, MIT, Cambridge, MA, 02139, USA}
\affil{Harvard-Smithsonian Center for Astrophysics, 60 Garden Street, Cambridge, MA, 02138, USA}

\author{Hans--Walter Rix}
\affil{Max Planck Institute f\"ur Astronomie, K\"onigstuhl 17, 69117 Heidelberg, Germany}

\author{Mark Swinbank}
\affil{Centre for Extragalactic Astronomy, Department of Physics, Durham University, South Road, Durham, DH1 3LE, UK}

\author{Mark Vogelsberger}
\affil{Kavli Institute for Astrophysics and Space Research, Department of Physics, MIT, Cambridge, MA, 02139, USA}

\author{Paul van der Werf}
\affil{Leiden Observatory, Leiden University, PO Box 9513, NL-2300 RA Leiden, The Netherlands}

\author{Jeff Wagg}
\affil{SKA Organization, Lower Withington Macclesfield, Cheshire SK11 9DL, UK}

\author{L. Y. Aaron Yung}
\affil{Department of Physics and Astronomy, Rutgers, The State University of New Jersey, 136 Frelinghuysen Rd,
Piscataway, NJ 08854, USA}

\begin{abstract}
{\gp 
The ALMA Spectroscopic Survey in the Hubble Ultra Deep Field (ASPECS)
provides new constraints for galaxy formation models on the molecular gas
properties of galaxies. We compare results from ASPECS to predictions from two cosmological galaxy
formation models: the IllustrisTNG hydrodynamical simulations and the Santa Cruz semi-analytic
model (SC SAM). We explore several recipes to model the \h2 content of
galaxies, finding them to be consistent
  with one another, and take into account the sensitivity limits and
  survey area of ASPECS. For a canonical CO--to--\h2 conversion factor of $\alpha_{\rm CO} =
3.6\,\rm{M}_\odot/(\rm{K}\,\rm{km/s}\,\rm{pc}^{2})$ the results of
our work include:  (1) the \h2 mass of $z>1$ galaxies predicted by the
models as a function of their stellar mass is a factor of 2--3 lower
than observed; (2) the models do not reproduce the
number of \h2-rich ($M_{\rm H2} > 3\times 10^{10}\,\rm{M}_\odot$)
galaxies observed by ASPECS; (3) the \h2 cosmic density evolution
predicted by  IllustrisTNG (the SC SAM) is in tension (only just agrees) with the observed cosmic density, even after accounting for the ASPECS selection function and field-to-field variance effects.
The tension between models and observations at $z>1$ can be alleviated by adopting a CO--to--\h2 conversion factor in the range $\alpha_{\rm CO} =
2.0 - 0.8\,\rm{M}_\odot/(\rm{K}\,\rm{km/s}\,\rm{pc}^{2})$. Additional
work on constraining the CO--to--\h2 conversion factor and CO
excitation conditions of galaxies through observations and theory will be necessary to more robustly test the success of
galaxy formation models.}
\end{abstract}



\section{Introduction}
Surveys of large fields in the sky have been instrumental for
our understanding of galaxy formation and evolution. A pioneering
survey was carried out with the {\it Hubble} Space Telescope
\citep[HST,][]{Williams1996}, pointing at a region in the sky now known
as the Hubble Deep Field (HDF). Ever since, large field surveys have been carried out at X-ray, optical, infrared, submillimeter (sub-mm)
continuum, and radio wavelengths. These efforts have revealed the
star-formation (SF) history of our Universe, quantified the stellar
build-up of galaxies, and have been used to derive
galaxy properties such as stellar masses, star-formation rates (SFR),
morphologies, and sizes over cosmic time \citep[e.g.,][]{Madau2014}. One of the most well known
results obtained is that the SF history of
our Universe peaked at redshifts $z\sim 2-3$, after which it dropped
to its present-day value \citep[e.g.,][for a
recent review see \citealt{Madau2014}]{Lilly1995,
Madau1996,Hopkins2004,Hopkins2006}.

Although the discussed efforts have shed light on the evolution of galaxy properties such as stellar mass, morphology, and SF, similar studies focusing on the gas
content, the fuel for star formation, have lagged behind. New and updated facilities operating in the millimeter and radio waveband such as the Atacama Large
(sub-)Millimeter Array (ALMA), NOrthern Extended Millimeter Array (NOEMA), and the Jansky Very Large Array (JVLA) have now made a survey of cold gas in our Universe feasible. A first
pilot to develop the necessary techniques was performed with the Plateau
de Bure Interferometer \citep{Decarli2014, Walter2014}. This was
followed by the first search for emission lines, mostly carbon monoxide ($^{12}$CO,
hereafter CO) using ALMA, focusing on a
small ($\sim$1 arcmin$^2$) region within the Hubble Ultra Deep Field
\citep[HUDF,][]{Walter2016, Decarli2016}. This effort is currently
extended (4.6 arcmin$^2$) as part
of `The ALMA Spectroscopic Survey in the Hubble Ultra Deep Field'
\citep[ASPECS,][]{Walter2016, Gonzalez2019, Decarli2019}. Among other goals, this survey aims to detect CO emission and fine-structure lines of carbon over cosmic time in the HUDF.  The CO emission
is used as a proxy for the molecular hydrogen gas content of galaxies (through a CO--to--\h2 molecular gas conversion factor). A complementary survey, COLDZ, has been carried out with the JVLA in GOODS-North and COSMOS \citep{Pavesi2018,Riechers2018}. The area covered on the sky by COLDZ is larger compared to ASPECS, but it is shallower (and focuses on CO J$=$1--0 instead of the higher rotational transitions targeted by ASPECS).

Surveys of a field on the sky are complementary to surveys targeting
galaxies based on some pre-selection. First of all, a survey without a pre-selection of targets 
allows one to detect classes of galaxies that would have potentially been missed
in targeted surveys because they do not fulfil the selection
criteria.  Second, these surveys are the perfect tool to measure the
number densities of different classes of galaxies. With this in mind,
one of the main science goals of ASPECS is to
quantify the \h2 mass function and \h2 cosmic density of the Universe over time. 

Surveys focusing on the gas content of galaxies and our Universe
provide an important constraint and additional challenge for theoretical models of galaxy
formation.  Theoretical models can be used to estimate limitations in the
observations (e.g., field-to-field variance, selection functions) and to put
the observational results into a broader context (gas baryon cycle,
galaxy evolution). On the other hand, observational constraints help
the modelers in better understanding the physics relevant for galaxy (and gas) evolution
(such as feedback and star-formation recipes), and they can serve as benchmarks to understand the strengths/limitations of models.

During the last decade a large number of groups have implemented the
modeling of \h2 in
post-processing or on-the-fly in hydrodynamic
\citep[e.g.,][]{Popping2009,Christensen2012,Kuhlen2012,Thompson2014,Lagos2015,Marinacci2017,
  Diemer2018, Stevens2018} and in (semi-)analytic models \citep[e.g.,][]{Obreschkow2009,Dutton2010,Fu2010,Lagos2011,Krumholz2012,Popping2014,Xie2017,Lagos2018}.
 Most of these models use metallicity- or pressure-based recipes to
separate the cold interstellar medium (ISM) into an atomic (\hi) and
molecular (\h2) component. The pressure-based recipe builds upon the
empirically determined relation between the mid-plane pressure acting
on a galaxy disc and the ratio between atomic and molecular hydrogen
\citep{Blitz2004,Blitz2006,Leroy2008}. The physical motivation for
the correlation between mid-plane pressure and molecular hydrogen mass
fraction was first presented in \citet{Elmegreen1989}. The
metallicity-based recipes (where the metallicity is a proxy for the dust grains that act as a catalyst for the formation of \h2) are often based on work presented in
\citet{Gnedin2011} or Krumholz and collaborators
\citep{Krumholz2008,Krumholz2009,Mckee2010,Krumholz2013}. \citet{Gnedin2011}
used high-resolution simulations including chemical networks to
derive fitting functions that relate the \h2 fraction of the ISM to the
gas surface density of galaxies on kpc scales, the metallicity, and
the strength of Ultraviolet (UV) radiation
field.  \citet{Krumholz2009} presented analytic models for the
formation of \h2 as a function of total gas density and metallicity,
supported by numerical simulations with simplified geometries
\citep{Krumholz2008,Krumholz2009}. This work was further developed in
\citet{Krumholz2013}.

In this paper we will compare predictions for the \h2 content of
galaxies by the IllustrisTNG (the next generation) model \citep{Weinberger2017,Pillepich2018} and the Santa Cruz semi-analytic model \citep[SC SAM,][]{Somerville1999, Somerville2001} to the results from the
ASPECS survey. We will specifically
try to quantify the success of these different galaxy formation
evolution models in reproducing the observations by accounting for sensitivity limits, field-to-field variance effects, and systematic theoretical uncertainties. We will furthermore
use these models to assess the importance of field-to-field variance and the
ASPECS selection functions on the conclusions drawn from the
survey. We encompass the
systematic uncertainties in the modeling of \h2 by employing three different prescriptions
to calculate the amount of molecular hydrogen.

IllustrisTNG is a cosmological, large-scale
gravity+magnetohydrodynamical simulation based on the moving mesh code
{\sc AREPO} \citep{Springel2010}. The SC SAM does not solve for the hydrodynamic equations,
but rather uses analytical recipes to describe the flow of baryons
between different `reservoirs' (hot gas, cold gas making up the
interstellar medium, ejected gas, and stars). Both models include prescriptions for physical processes such as the cooling and accretion of gas onto galaxies, star-formation, stellar and black
hole feedback, chemical enrichment, and stellar evolution.

Although these two
models are different in nature and have different  strengths and disadvantages, they both reasonably reproduce
some of the key observables of the galaxy population in our local
Universe, such as the galaxy stellar mass function, sizes, and SFR of
galaxies (at least at low redshifts). The different nature of these two models probes
the systematic uncertainty across models when these are used to interpret
observations. Furthermore, any shared successes or problems of these
two models may point to a general success/misunderstanding of galaxy
formation theory rather than model dependent uncertainties.

This paper is organised as follows. In Section \ref{sec:models}
we briefly present IllustrisTNG, the SC SAM, and the implementation of the various \h2 recipes. We provide a brief overview of ASPECS in Section \ref{sec:ASPECS_overview}. In
Section \ref{sec:results} we present the predictions by the different
models and how these compare to the results from ASPECS. We
discuss our results in Section \ref{sec:discussion} and present a
short summary and our conclusions in Section \ref{sec:conclusions}. Throughout this paper we assume a Chabrier stellar initial mass function
\citep[IMF;][]{Chabrier2003} in the mass range 0.1--100
$\rm{M}_\odot$ and adopt a cosmology consistent with the recent Planck
results \citep[$\Omega_{m} = 0.31,\,\Omega_{\Lambda} =
0.69,\,\Omega_{b} = 0.0486,\,h = 0.677,\,\sigma_8 = 0.8159,\,n_s = 0.97$]{Planck2016}. All presented gas masses (model predictions and observations) are pure hydrogen masses (do not include a correction for helium).

\section{Description of the models}
\label{sec:models}
\subsection{IllustrisTNG}
\label{sec:IllustrisTNG}
In this paper we use and analyze the TNG100 simulation, a
  $\sim$(100 Mpc$)^3$ cosmological volume simulated with the code {\sc
    AREPO} \citep{Springel2010} within the IllustrisTNG
  project\footnote{www.tng-project.org} \citep{Pillepich2018b,
    Naiman2018, Nelson2018, Springel2018,Marinacci2018}. The
  IllustrisTNG model is a revised version of the Illustris galaxy
  formation model \citep{Vogelsberger2013, Torrey2014}. TNG100 evolves cold dark matter (DM) and gas from early times to $z=0$ by solving for the coupled equations of gravity and magneto-hydrodynamics (MHD) in an expanding Universe (in a standard cosmological scenario, \citealt{Planck2016}) while including prescriptions for star formation, stellar evolution and hence mass and metal return from stars to the interstellar medium (ISM), gas cooling and heating, feedback from stars and feedback from supermassive black holes \citep[see][for details on the IllustrisTNG model]{Weinberger2017, Pillepich2018}. 

At $z=0$, TNG100 samples many thousands of galaxies above $M_* \simeq
10^{10}\,\rm{M}_\odot$ in a variety of environments, including for
example ten massive clusters above $M \simeq 10^{14}\,\rm{M}_\odot$
(total mass). The mass resolution of the simulation is uniform across
the simulated volume (about $7.5\times 10^6\rm{M}_\odot$  for DM
particles and $1.4\times 10^6\,\rm{M}_\odot$ for both gas cells and
stellar particles). The gravitational forces are softened for the
collisionless components (DM and stars) at about 700 pc at $z=0$,
while the gravitational softening of the gas elements is adaptive and can be as small
as $\sim280$ pc. The spatial resolution of the hydrodynamics is fully
adaptive, with smaller gas cells at progressively higher densities: in
the star forming regions of galaxies, the average gas-cell size in
TNG100 is about 355 pc \citep[see table A1 in][for more details]{Nelson2018}. 

The TNG100 box (or TNG, for brevity, throughout this paper) is a rerun of the original Illustris simulation \citep{Vogelsberger2014a,Vogelsberger2014b, Genel2014, Sijacki2015} with updated and new aspects of the galaxy-physics model, including -- among others -- MHD, modified galactic winds, and a new kinetic, black hole-driven wind feedback model. Importantly for this paper, in the Illustris and IllustrisTNG frameworks, gas is converted stochastically into stellar particles following the two-phase ISM model of \citet{Springel2003}: when a gas cell exceeds a density threshold ($n_{\rm H} \simeq 0.1$cm$^{-3}$), it is dubbed {\it star forming}, irrespective of its metallicity. This model prescribes that low-temperature and high-density gas (below about $10^4$ K and above the star-formation density threshold) is placed on an equation of state between e.g. temperature and density, meaning that the multi-phase nature of the ISM at higher densities (or colder temperatures) is assumed, rather than hydrodynamically resolved. In these simulations, the production and distribution of nine chemical elements is followed (H, He, C, N, O, Ne, Mg, Si, and  Fe) but no distinction is made between atomic and molecular phases, which hence need to be modeled in post processing for the purposes of this analysis (see subsequent sections). Gas radiatively cools in the presence of a spatially uniform, redshift-dependent, ionizing UV background radiation field \citep{Faucher2009}, including corrections for self-shielding in the dense ISM but neglecting local sources of radiation. Metal-line cooling and the effects of a radiative feedback from supermassive black holes are also taken into account in addition to energy losses induced by two-body processes (collisional excitation, collisional ionization, recombination, dielectric recombination and free-free emission) and inverse Compton cooling off the CMB. 

While a certain degree of freedom is unavoidable in these models (mostly owing to the subgrid nature of a subset of the physical ingredients), their parameters are chosen to obtain a reasonable match to a small set of observational, galaxy-statistics results. For IllustrisTNG, these chiefly included the current baryonic mass content of galaxies and haloes and the galaxy stellar mass function at $z=0$ \citep[see][for details]{Pillepich2018}. The IllustrisTNG outcome is consistent with a series of other observations, including the galaxy stellar mass functions at $z\lesssim4$ \citep{Pillepich2018b}, the galaxy color bimodality observed in the Sloan Digital Sky Survey \citep{Nelson2018}, the large-scale spatial clustering of galaxies also when split by galaxy colors \citep{Springel2018}, the gas-phase oxygen abundance and distribution within \citep{Torrey2018} and around galaxies \citep{Nelson2018b}, the metallicity content of the intra-cluster medium \citep{Vogelsberger2018},  
and the average trends, evolution, and scatter of the galaxy stellar
size-mass relation at $z\lesssim2$ \citep{Genel2018}. Thanks to such
general validations of the model, we can use the IllustrisTNG galaxy
population as a plausible synthetic dataset for further studies,
particularly at the intermediate and high redshifts that are probed by
ASPECS and that had not been considered for the model development (the gas mass fraction within galaxies were not used to constrain the model, particularly at high redshifts, which makes the current exploration interesting).

\subsubsection{Input parameters for \h2 recipes in IllustrisTNG}
In order to obtain the molecular gas content of simulated galaxies (see Section \ref{sec:H2_recipes}), we employ a number of approaches to calculate the molecular hydrogen
fraction $f_{\rm H2}$ ($=M_{\rm H2}/M_{\rm Hydrogen}$) of gas cells
within the simulation. The gas
cells represent a mixture of Hydrogen, Helium, and metals. {\gp Although in the TNG calculations the fraction of hydrogen is tracked on a gas cell by cell basis, this is not always stored in the output data. Namely, the Hydrogen fraction is stored only in 20 of 100 snapshots (in the so-called full snapshots) and not for all the redshifts we intend to study. For these reasons, we simply assume a hydrogen fraction for the gas cells of $f_{\rm H} = 0.76$.}

\paragraph{Gas surface density} Some of the recipes employed to compute the molecular hydrogen fraction of the cold gas depend on the cold gas surface density. To calculate the gas surface density of a gas
cell we multiply its gas density with the characteristic Jeans length
belonging to that cell \citep[following e.g.,][]{Lagos2015,Marinacci2017}. The Jeans
  length $\lambda_{\rm J}$ is calculated as
\begin{equation}
\lambda_{\rm J} = \sqrt\frac{c^2_{\rm s}}{G\rho} =
\sqrt\frac{\gamma(\gamma - 1)u}{G\rho},
\end{equation}
where $c_{\rm s}$ is the sound speed of the gas, $G$ and $\rho$
represent the gravitational constant and total gas density of  a cell,
respectively, $u$ the internal energy of the gas cell, and $\gamma
= 5/3$ the ratio of heat capacities. In the case of star-forming cells the internal energy represents
  a mix between the hot ISM and star-forming gas. For these cells we
  take the internal energy to be $T_{\rm SF} = 1000\,\rm{K}$ \citep{Springel2003,Marinacci2017}.

The hydrogen gas surface density of each cell is then calculated
  as
\begin{equation}
\Sigma_{\rm H} = f_{\rm H}f_{\rm neutral,H}\lambda_{\rm J}\rho,
\end{equation}
where $f_{\rm neutral,H}$ marks the {\gp fraction of hydrogen in a
gas cell that is neutral} (i.e., atomic or molecular). We assume  $f_{\rm
  neutral,H}=1.0$ for star-forming cells, whereas we adopt the value
suggested from IllustrisTNG for  $f_{\rm neutral,H}$ for non
star-forming cells. 

\paragraph{Radiation field}
For a subset of the employed recipes the molecular hydrogen fraction also
depends on the local UV radiation field $G_0$. The local UV radiation field $G_0$ impinging on the gas cells is
  calculated differently for star-forming and non-star-forming
  cells. For star-forming cells we scale $G_0$ with the local SFR surface density ($\Sigma_{\rm SFR}$, calculated by multiplying the star-formation rate density of each cell by the
Jeans length) such that
\begin{equation}
G_0= \frac{\Sigma_{\rm SFR}}{\Sigma_{\rm SFR, MW}},
\end{equation}
where $\Sigma_{\rm SFR,MW} = 0.004\,\rm{M}\,\rm{yr}^{-1}\,\rm{kpc}^{-2}$
is the local SFR surface density in the MW \citep{Robertson2008}. We
note that the local value for the MW SFR surface density is
somewhat uncertain, varying in the range (1 -- 7) $\times
10^{-3}\,\rm{M}_\odot\,\rm{yr}^{-1}\,\rm{kpc}^{-2}$ \citep{Miller1979,
  Bonatto2011}. We scale the UV radiation field for non-star-forming
cells  as a function of the time-dependent \hi heating rate from
\citet{Faucher2009} at 1000 \AA. \citet{Diemer2018} adopted a
different approach to calculate the UV radiation field impinging on
every gas cell by propagating the UV radiation from star-forming
particles to its surroundings, accounting for dust absorption.  The
median difference in the predicted \h2 mass by Diemer et al. and our
method is 15 \% for galaxies with \h2 masses more massive than
$10^{9}\,\rm{M}_\odot$ at the redshifts that are relevant for ASPECS
(at $z=0$ this is $\sim 40$ \% for the GK method).

\paragraph{Dust}  
The dust abundance of the cold gas in terms of the MW dust
  abundance $D_{\rm MW}$ is assumed to be equal to the gas-phase
  metallicity expressed in solar units, i.e., $D_{\rm MW} = Z/Z_\odot$. Both
  observations and simulations have demonstrated that this scaling is
  appropriate over a large range of gas-phase metallicities
  \citep[$Z\geq 0.1\,Z_\odot$,][]{Remy-Ruyer2014,McKinnon2017,Popping2017dust}.

\subsection{Santa Cruz semi-analytic model}
\label{sec:SCSAM}
The SC semi-analytic galaxy formation model was first presented in
\citet{Somerville1999} and \citet{Somerville2001}. Updates
to this model were described in \citet[][S08]{Somerville2008},
\citet{Somerville2012}, \citet[PST14]{Popping2014},
\citet{Porter2014}, and \citet[SPT15]{Somerville2015}. The model
tracks the hierarchical clustering of dark matter haloes, shock
heating and radiative cooling of gas, SN feedback, star formation,
active galactic nuclei (AGN) feedback (by quasars and radio jets),
metal enrichment of the interstellar and intracluster media, disk
instabilities, mergers
of galaxies, starbursts, and the evolution of stellar
populations. PST14 and SPT15 included new recipes that track the
amount of ionized, atomic, and molecular hydrogen in galaxies and included a
molecular hydrogen based star-formation recipe. The SC SAM
has been fairly successful in
reproducing the local properties of galaxies such as the stellar mass
function, gas fractions, gas mass function, SFRs, and stellar
metallicities, as well as the evolution of the galaxy sizes, quenched
fractions, stellar mass functions, dust content, and luminosity functions
\citep[PST14,
SPT15]{Somerville2008,Somerville2012,Porter2014,Popping2014PDR,Brennan2015,Popping2016,
Popping2017dust,Yung2018}.

The semi-analytic framework essentially describes the flow of material between different types of reservoirs. All galaxies form within
a dark matter halo. There are three reservoirs for gas; the ``hot'' gas
that is assumed to be in a quasi-hydrostatic spherical configuration
throughout the virial radius of the halo; the ``cold'' gas in the
galaxy, assumed to be in a thin disk; and the ``ejected'' gas which is gas
 that has been heated and ejected from the halo by stellar winds. 
Differential equations describe the movement of gas between these three reservoirs. As dark matter halos grow in mass, pristine gas is accreted
from the intergalactic medium into the hot halo. A cooling model is used to calculate the rate at which gas accretes from
the hot halo into the cold gas reservoir, where it becomes available
to form stars. Gas participating in star formation is removed from the
cold gas reservoir and locked up in stars. Gas can furthermore be
removed from the cold gas reservoir by stellar and AGN-driven
winds. Part of the gas that is ejected by stellar winds is returned to
the hot halo, whereas the rest is deposited in the ``ejected''
reservoir. The fraction of gas that escapes the hot halo
  is calculated as a function of the virial velocity of the progenitor galaxy (see
  S08 for more detail). Gas ``re-accretes'' from the ejected
reservoir back into the hot halo according to a parameterized
timescale (again see S08 for details). 

The galaxy that initially forms at the center of each halo is called
the ``central'' galaxy. When dark matter halos merge, the central
galaxies in the smaller halos become ``satellite'' galaxies. These
satellite galaxies orbit
within the larger halo until their orbit decays and they merge with
the central galaxy, or until they are tidally destroyed.

We make use of merger trees extracted from the Bolshoi N-body dark
matter simulation
\citep{Klypin2011,Trujillo-Gomez2011,Rodriguez-Puebla2016}, using a
box with a size of 142 cMpc on each side {\gp (which is a subset of the
  total Bolshoi simulation, which spans $\sim$370 cMpc on each side)}. Dark
matter haloes were identified using the {\sc ROCKSTAR} algorithm \citep{BehrooziRockstar2013}. This simulation
is complete down to haloes with a mass of $M_{\rm vir} = 2.13 \times 10^{10}\,\rm{M}_\odot$, with a force resolution of 1  kpc $h^{-1}$ and a mass resolution of $1.9 \times 10^8\,\rm{M}_\odot$
per particle. The model parameters adopted in this work are the same as in SPT15, except for $\alpha_{\rm rh} = 2.6$ (the slope
of the SN feedback strength as a function of galaxy circular velocity)
and $\kappa_{\rm AGN} =3.0\times10^{-3}$ (the strength of the radio
mode feedback). These parameters were set by calibrating the model to
the redshift zero stellar mass -- halo mass relation, the $z=0$
stellar mass function,  the $z=0$
stellar mass--metallicity relation, the $z=0$ total cold gas fraction
(\hi $+$ \h2) of galaxies, and the black hole -- bulge mass
relation. Like IllustrisTNG, we did not use $z>0$ gas masses as
constraints when calibrating the SC SAM. More details on the free parameters can be found in
S08 and SPT15.

\subsubsection{Input properties for molecular hydrogen recipes in the SC SAM}
We assume that the cold gas (\hi $+$ \h2) is distributed in an
exponential disc with scale radius $r_{\mathrm{gas}}$ with a central gas surface density of $m_{\rm cold}/(2\pi\,r_{\mathrm{gas}}^2)$, where $m_{\rm cold}$ is the
mass of all cold gas in the disc. This is a good approximation for
nearby spiral galaxies \citep{Bigiel:2012}. The stellar scale
length is defined as $r_{\mathrm{star}} =
r_{\mathrm{gas}}/\chi_{\mathrm{gas}}$, with $\chi_{\mathrm{gas}}=1.7$
fixed to match stellar scale lengths at $z = 0$. The gas
disc is divided into radial annuli and the fraction of molecular gas within
each annulus is calculated as described below. The integrated mass of \hi and \h2 in
the disc at each time step is calculated using a fifth order
Runga-Kutta integration scheme.

The cold gas consists of an ionized, atomic and
molecular component. The radiation field from stars within the galaxy and an external background are responsible for the ionized component. The fraction of gas ionized by the stars in the galaxy is described as 
$f_{\rm ion, int}$. The
external background ionizes a slab of gas on each side of the
disc. Assuming that all the
gas with a surface density below some critical value $\Sigma_{\rm
  HII}$ is ionized, we  use \citep{Gnedin2012}
{\gp 
\begin{equation}
 f_{\rm ion,bg} = \frac{\Sigma_{\rm HII}}{\Sigma_0}
\left[1 + \ln \left(\frac{\Sigma_0}{\Sigma_{\rm HII}} \right) + 0.5
  \left(\ln \left(\frac{\Sigma_0}{\Sigma_{\rm HII}}\right) \right)^2
  \right]
\end{equation} to described the fraction of gas ionized by the UV background.
The total ionized fraction can then be expressed as $f_{\rm ion} =
f_{\rm ion, int} + f_{\rm ion, bg}$.}
Throughout this paper we assume $f_{\rm ion, int} = 0.2$ (as in the
Milky Way) and $\Sigma_{\rm HII} = 0.4 \, M_\odot \rm{pc}^{-2}$,
supported by the results of \citet{Gnedin2012}.

\subsection{Molecular hydrogen fraction recipes}
\label{sec:H2_recipes}
In this paper we present predictions for the \h2 properties of galaxies by adopting three different molecular hydrogen fraction recipes. The first is a metallicity-based recipe based on work by \citet[GK]{Gnedin2011}, the second a metallicity-based recipe from \citet[K13]{Krumholz2013}, and the last an empirically derived recipe based on the mid-plane pressure acting on the disk of galaxies \citep[BR]{Blitz2006}. In most of this paper (except for Section \ref{sec:inherent}) we only show the predictions for the GK recipe. In the current section we present the GK recipe, whereas the BR and K13 recipes are described in detail in the appendix of this work.

\subsubsection{Gnedin \& Kravtsov 2011 (GK)}
\label{sec:GK}
The first \h2 used in this work is based on the work by \citet{Gnedin2011} to compute
the \h2 fraction of the cold gas. The authors performed detailed simulations including non-equilibrium chemistry and simplified 3D on-the-fly radiative transfer calculations. Motivated by their simulation results, the authors present fitting formulae for the \h2 fraction of cold gas. The \h2 fraction depends on the 
dust-to-gas ratio relative to solar, $D_{\rm MW}$, the ionising background radiation
field, $G_0$, and the surface density of the cold gas,
 $\Sigma_{\rm HI + H2}$. 
The molecular hydrogen fraction of the cold gas is given as
\begin{equation}
 f_{H_2} = \left[1+\frac{\tilde{\Sigma}}{\Sigma_{HI+H_2}}\right]^{-2} 
\end{equation}
where
\begin{eqnarray*}
\tilde{\Sigma}  & = &  20\, {\rm M_\odot pc^{-2}} \frac{\Lambda^{4/7}}{D_{\rm MW}} 
\frac{1}{\sqrt{1+G_0 D_{\rm MW}^2}}, \\
\Lambda & = & \ln(1+g D_{\rm MW}^{3/7}(G_0/15)^{4/7}),\\
g & = & \frac{1+\alpha s + s^2}{1+s},\\
s &  = & \frac{0.04}{D_*+D_{\rm MW}},\\
\alpha &  = & 5 \frac{G_0/2}{1+(G_0/2)^2},\\
D_* & = & 1.5 \times 10^{-3} \, \ln(1+(3G_0)^{1.7}).
\end{eqnarray*}

\subsubsection{The \h2 mass of a galaxy in IllustrisTNG}
\label{sec:aperture}
Individual galaxies within IllustrisTNG and their properties
correspond to sub-haloes within the IllustrisTNG volume. One
measurement of the gas mass of a subhalo is the sum over all gas cells
gravitationally bound to it. This gas mass does not
necessarily correspond to the gas mass that observations
would probe. In most of this paper we will use two operational definitions
for the \h2 mass of galaxies. The first includes the \h2 mass of all
the cells that are gravitationally bound to the subhalo (`Grav'). The second only accounts for
the \h2 mass of cells that are within a circular aperture with a
diameter corresponding to  3.5
arcsec on the sky, centered around the galaxy (`3.5arcsec'). This
aperture has the same size as the beam of the cube from which the flux
of galaxies in the ASPECS survey is extracted (see next Section). At a redshift of
exactly $z=0$ such a beam
corresponds to a infinitesimal area on the sky. We thus replace the
`3.5arcsec' aperture at $z=0$ by an aperture corresponding to two
times the stellar half-mass radius of the galaxy (`In2Rad'). This is a
closer (but not perfect) match to the observations used to control the validity of the
model at $z=0$ (\citealt{Diemer2019} presents a robust comparison between model predictions and observations at $z=0$, better accounting for aperture variations between different observations at $z=0$). By definition the \h2 masses predicted by the SAM
correspond to the `Grav' aperture for IllustrisTNG.

\subsubsection{Metallicity and molecular hydrogen fraction floor in the SC SAM}
Following PST14 and SPT15, we adopt a metallicity floor of
$Z=10^{-3}\,$Z$_\odot$ and a floor for the fraction of molecular
hydrogen of $f_{\rm mol}=10^{-4}$.  These floors represent the
enrichment of the ISM by `Pop III' stars and the formation of
molecular hydrogen through channels other than on dust grains
\citep{Haiman1996,Bromm2004}. SPT15 showed that the SC semi-analytic model results are not
  sensitive to the precise values of these parameters.

\section{ASPECS survey overview}
\label{sec:ASPECS_overview}
We compare our models and predictions with the observational results from molecular field campaigns. The ALMA Spectroscopic Survey in the Hubble Ultra Deep Field (ASPECS LP) is an ALMA Large Program (Program ID: 2016.1.00324.L) which consists of two scans, at 3\,mm and 1.2\,mm. The survey builds on the experience of the ASPECS Pilot program \citep{Walter2016,Aravena2016,Decarli2016}.  The 3\,mm campaign discussed here scanned a contiguous area of $\sim4.6$\,arcmin$^2$ in the frequency range 84--115\,GHz (presented in \citealt{Gonzalez2019} and \citealt{Decarli2019}). The targeted area matches the deepest HST near-infrared pointing in the HUDF. The frequency scan provides CO coverage at $z<0.37$, $1.01<z<1.74$, and at any $z>2.01$ (depending on CO transitions), thus allowing us to trace the evolution of the molecular gas mass functions and of $\rho$(H$_2$) as a function of redshift. 

{\gp The ASPECS LP reached a 5-$\sigma$ luminosity floor (i.e. brighter
sources correspond to a higher than 5-$\sigma$ certainty)},  of $\sim 2\times10^9$\,K\,km\,s$^{-1}$\,pc$^2$ (assuming a line width of 200\,km\,s$^{-2}$) at virtually any redshift $z>1$, and encompassed a volume of 338\,Mpc$^3$, 8198\,Mpc$^3$,
14931\,Mpc$^3$, 18242\,Mpc$^3$, in CO(1-0), CO(2-1), CO(3-2), CO(4-3),
respectively. The line-search is performed in a cube with a
synthesized beam of $\approx 1.75'' \times 1.49''$. Once lines are
detected, their spectra are extracted from a cube for which the
angular resolution is lowered to a beam size of $\sim$3.5'', in order to capture all
the emission that would have been resolved in the original cube. The
lines used in the construction of the luminosity functions are
identified exclusively based on the ASPECS LP 3\,mm
dataset, with no support from prior information from catalogs built at
other wavelengths. This allows us to circumvent any selection bias in
the targeted galaxies, thus providing a direct census of the gas
content in high-redshift galaxies. The line search resulted in 16
lines detected at S/N$>$6.4 {\gp (i.e., the sources with a fidelity of
  100\%, we refer the reader to
\citealt{Gonzalez2019} and \citealt{Boogaard2019} for a more detailed discussion on the detected
lines, their S/N ratio, fidelity, and the fraction of galaxies that were recovered
in the Hubble Ultra Deep Field)}. The
impact of false positive detections and the completeness of our search
are discussed in \citet{Gonzalez2019}. The lines are then identified
by matching the discovered lines with the rich multi-wavelength legacy
dataset collected in the HUDF, and in particular the redshift catalog
provided by the MUSE HUDF survey (\citealt{Bacon2017},
\citealt{Inami2017}). When a counterpart is found, we refer to its
spectroscopic or photometric redshift to guide the line identification
(and thus the redshift measurement); otherwise, we assign the redshift
based on a Monte Carlo process. Details of this analysis are presented in \citet{Decarli2019} and \citet{Boogaard2019}. The line luminosities are then transformed into corresponding CO(1-0) luminosities based on the \citet{Daddi2015} CO SLED template, which is intermediate between the case of low excitation (as in the Milky Way) and a thermalized case \citep[see, e.g.,][]{Carilli2013}. Finally, CO(1-0) luminosities are converted into molecular gas masses based on a fixed $\alpha_{\rm CO}=3.6$ M$_\odot$\,(K\,km\,s$^{-1}$\,pc$^2$)$^{-1}$ \citep[following, e.g.,][]{Decarli2016}. The choice of a relatively high $\alpha_{\rm CO}$ is justified by the finding of solar metallicity value for all the detected galaxies in our field for which metallicity estimates are available \citep{Boogaard2019}. The molecular gas mass can easily be rescaled to different assumptions for these conversion factors following : M$_{\rm H2}$ / M$_\odot$\,=\,($\alpha_{\rm CO}/\rm{r_{J1}}) \times L'_{\rm CO(J \rightarrow J-1)}$/(K km s$^{-1}$ pc$^2$)${−1}$, where $\rm{r}_{\rm J1}$ marks the ratio between between the CO J$=$1--0 and higher order rotational J transition luminosities, and $L'_{\rm CO(J \rightarrow J-1)}$ the observed CO $(J \rightarrow J-1)$ line luminosity in (K km s$^{-1}$ pc$^2$). The typical gaseous reservoirs identified in ASPECS have masses of $M_{\rm H2}=0.5-10\times10^{10}$\,M$_\odot$. 

\begin{figure*}
\centering
\includegraphics[scale = 0.85]{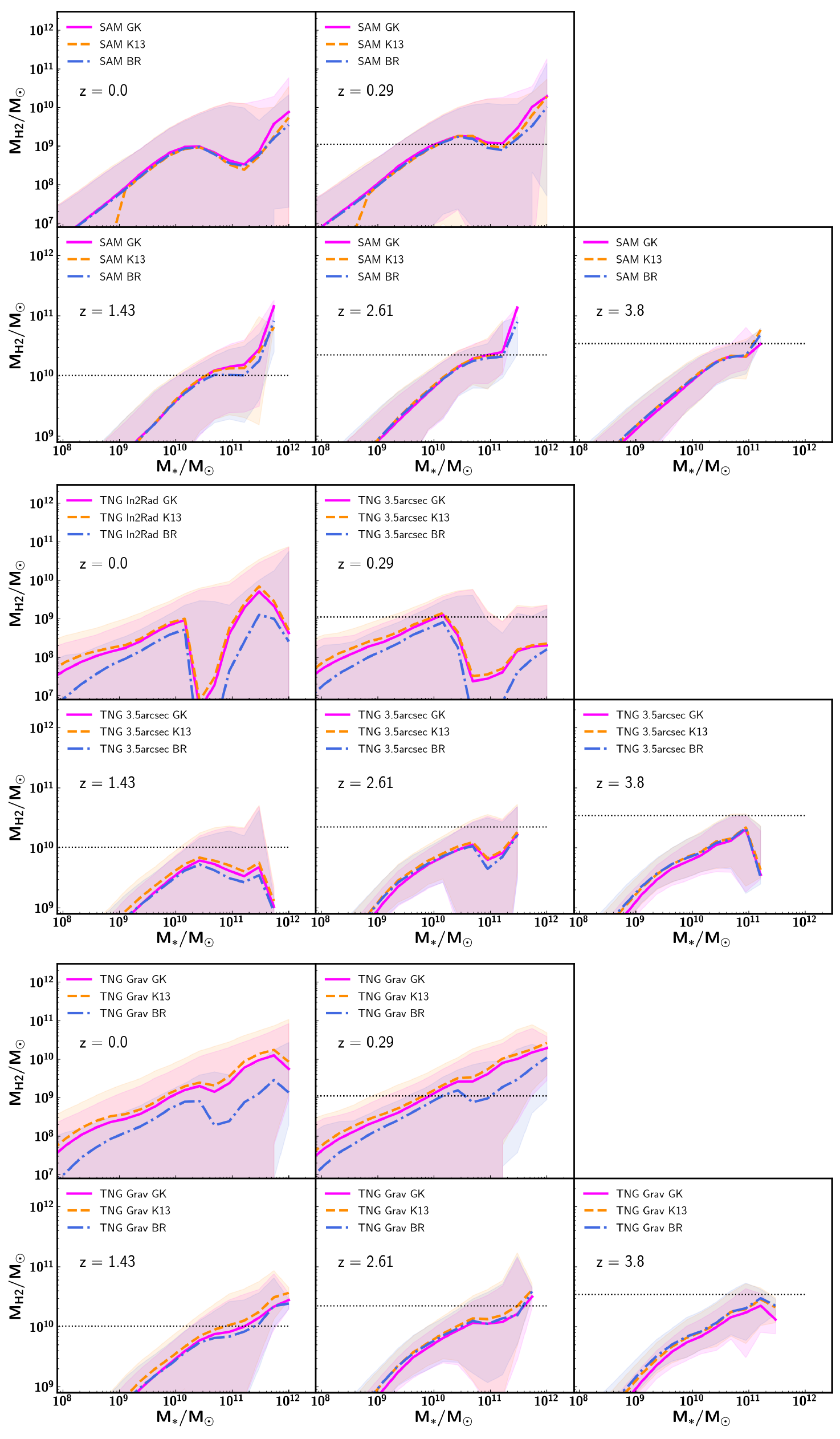}
\caption{{\gp The \h2 mass of galaxies at different redshifts as a function of their
  stellar mass as predicted by the models. No galaxy selections were applied
  to the model galaxy population. The top two rows correspond to the SC
  SAM. The middle two rows depict IllustrisTNG when adopting
  the `3.5arcsec' aperture ( note that at $z=0$ we use the `In2Rad'
  aperture). The bottom two rows show IllustrisTNG when
  adopting the `Grav' aperture.} In all cases, we show results with
the three \h2 partitioning recipes adopted in this work (GK: solid
pink; K13: dashed orange; BR: dotted-dashed blue). The thick lines
mark the median of the galaxy population, whereas the shaded regions
{\gp mark the two-sigma scatter} of the population. The dotted black horizontal line marks the sensitivity limit of ASPECS. \label{fig:mstar_vs_H2_nocut}}
\end{figure*}

\begin{figure*}
\includegraphics[width = \hsize]{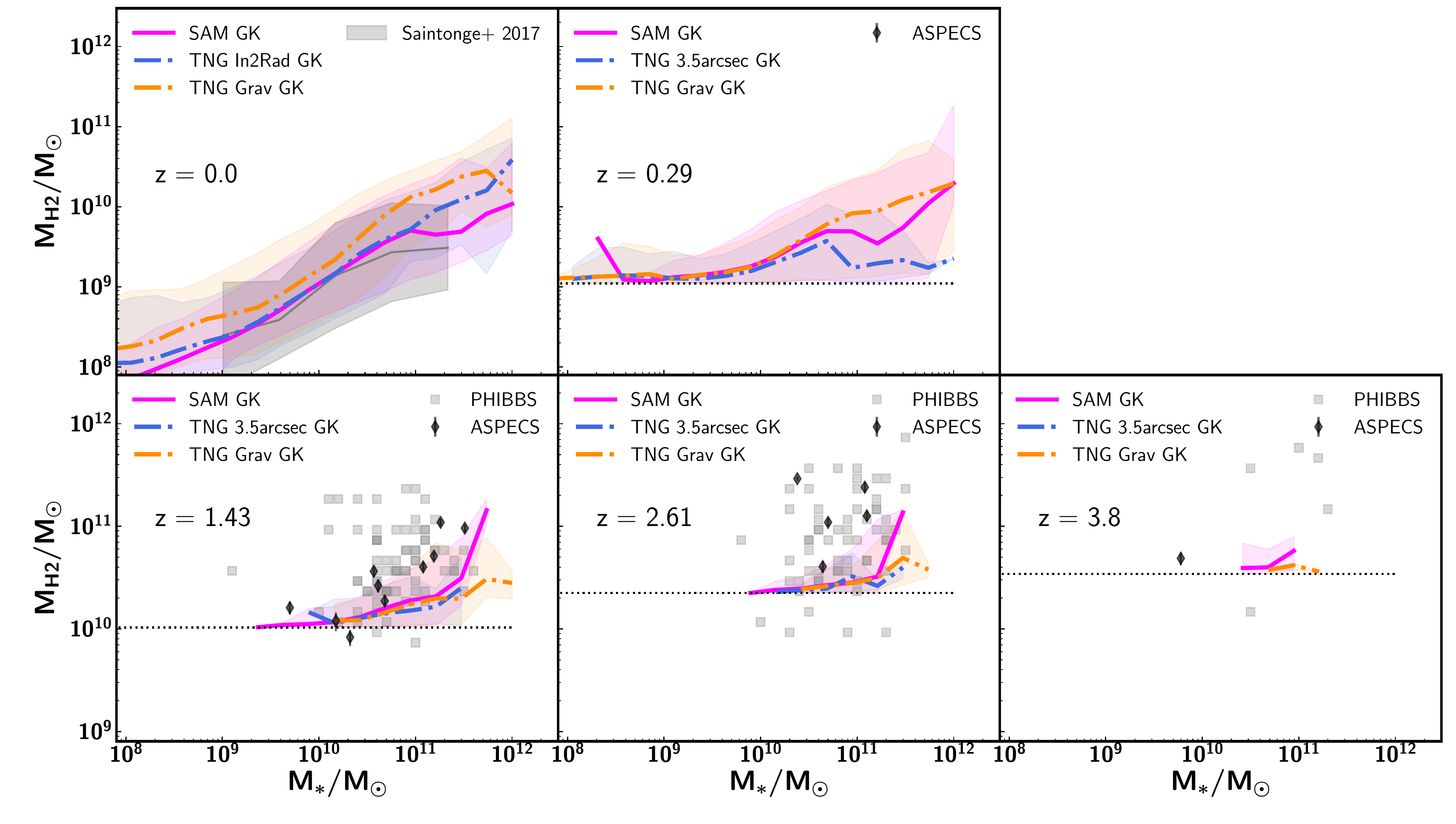}
\caption{The predicted and observed \h2 mass of galaxies at different redshifts as a function of their stellar mass. For the theoretical data, we account for observational selection effects. The results from SC SAM (solid pink) and from IllustrisTNG are shown by adopting the GK \h2 partitioning recipe. We show predictions for IllustrisTNG when
  adopting the `3.5arcsec' (dashed blue) and `Grav' (dotted-dashed orange) apertures (at $z=0$, the
  `3.5arcsec' aperture is replaced by the `In2Rad' aperture). In this Figure we assume $\alpha_{\rm CO} =
3.6\,\rm{M}_\odot/(\rm{K}\,\rm{km/s}\,\rm{pc}^{2})$. At $z=0$ a comparison is done to
  observational data from \citet{Saintonge2017}. To allow for a fair
  comparison and remove the contribution by quiescent galaxies, a selection criterion of $\log{\rm{SFR}} >
  \log{\rm{SFR}_{\rm MS}}(\rm{M}_*) - 0.4$ is applied to both the
  observed and modeled galaxies at $z=0$, where $\log{\rm{SFR}_{\rm
      MS}}(\rm{M}_*) $ marks the SFR of galaxies on the main sequence
  of star formation, following the definition of
  \citet{Speagle2014}. At higher redshifts model predictions are
  compared to the detections of ASPECS, as well as
  the compilation of CO detected galaxies presented as a part of PHIBBS in
  \citet{Tacconi2018}.  At these redshifts the ASPECS selection
  function is applied to the model galaxies (and depicted by the
  dotted black horizontal line). The solid lines mark the median of
  the galaxy population, whereas the shaded regions mark the
  {\gp two-sigma scatter} of the population. The different models are only partially able to reproduce the
  ASPECS and PHIBBS detections. We furthermore find that the ASPECS
  sensitivity sets a strong cut on the overall galaxy population
  (compare to Figure \ref{fig:mstar_vs_H2_nocut}).
\label{fig:scale_evol_SAMvsTNG}}
\end{figure*}

\section{Results and comparison to observations}
\label{sec:results}
In this Section we compare the \h2 model predictions by the IllustrisTNG simulation and
the SC SAM to the results of the ASPECS
survey, by adopting a CO--to--\h2 conversion factor of $\alpha_{\rm CO} =
3.6\,\rm{M}_\odot/(\rm{K}\,\rm{km/s}\,\rm{pc}^{2})$ for the observations following the ASPECS survey (we will change this assumption in our discussion in Section \ref{sec:discussion}). Where appropriate, we also include additional datasets to allow
for a broader comparison and to take into account observational sensitivity limits and field-to-field variance effects.

\subsection{\h2 scaling relation}
\label{sec:scale_evol}

\subsubsection{Inherent results}
\label{sec:inherent}
We present the \h2 mass of galaxies predicted from IllustrisTNG and the
SC SAM as a function of their stellar mass at $z=0$ and the median
redshifts of ASPECS in Figure \ref{fig:mstar_vs_H2_nocut}. This figure
includes all modeled galaxies at a redshift (i.e., no selection
function is applied) and shows predictions for the \h2 mass based on
all \h2 partitioning recipes considered in this work. We show the
predictions for IllustrisTNG when adopting the `Grav' aperture and the
`3.5arcsec' aperture (at $z=0$ replaced by the `In2Rad' aperture). We depict for reference the sensitivity limit of ASPECS as a dotted horizontal line in all the panels corresponding to galaxies at $z>0$ \citep[adopting the same CO excitation conditions and CO--to--\h2
  conversion factor as ASPECS, $\alpha_{\rm CO} =
3.6\,\rm{M}_\odot/(\rm{K}\,\rm{km/s}\,\rm{pc}^{2})$, and assume a CO
line width of 200 km/s {\gp (a typical value
  for main-sequence galaxies at $z>1$; a narrower linewidth yields a lower
mass limit, whereas a broader linewidth yields a higher mass limit,
see \citealt{Gonzalez2019} and Figure 9 in \citealt{Boogaard2019} for a detailed discussion on the effect of
the CO line width on the recovering fraction of galaxies and the \h2 sensitivity limit)}, for a detailed explanation of these choices see Section \ref{sec:ASPECS_overview} and][]{Decarli2019}. 

 Firstly, we find no significant difference in the predicted
average \h2 mass of galaxies by the three different \h2 partitioning recipes coupled to the SC SAM. When coupled to IllustrisTNG the GK and K13 recipes yield almost identical results. This is in line with the broader findings by \citealt{Diemer2018}. The BR recipe predicts lower \h2 masses at $z < 0.3$, but identical \h2 masses at higher redshifts. Given the minimal deviations in the medians between the different \h2 partitioning recipes, we will show from now on only predictions by the GK partitioning method in the main body of this paper. Model predictions obtained when adopting the other \h2 partitioning recipes are provided in Appendix \ref{sec:partitioning_recipes}.

Importantly, we find the \h2 mass of galaxies to increase as a
function of stellar mass for the SC SAM and IllustrisTNG when adopting
the `Grav' aperture, independent of redshift. At $z<3$ we see a decrease in the median \h2
mass for galaxies with a stellar mass larger than
$10^{10}\,\rm{M}_\odot$. This decrease is stronger for the SC SAM than
for IllustrisTNG with the `Grav' aperture. This drop in the median represents the
contribution from passive galaxies that host little molecular
hydrogen, driven by the active galactic nucleus feedback mechanism. These galaxies have \h2 masses that are below the
sensitivity limit of ASPECS. The upturn at the highest stellar masses corresponds to
a low number of central galaxies that are still relatively gas rich.

When adopting the `3.5arcsec' aperture for IllustrisTNG we see a different behaviour
from the `Grav' aperture. At $z=0.29$ and $z=1.43$ there is a much stronger
drop in the median \h2 mass of galaxies at masses larger than
$10^{10}\,\rm{M}_\odot$. This suggests that the bulk of the \h2
reservoir of the subhalos is outside of the aperture corresponding to
the ASPECS beam at these redshifts. A beam with a diameter of 3.5 arcsec at
$z=0.29$ corresponds to a size smaller than 2 times the stellar
half-mass radius of the galaxies in IllustrisTNG with $M_* >
10^{10}\,\rm{M}_\odot$ \citep{Genel2018}, suggesting that not all the
molecular gas close to the stellar disk is
captured. An AGN may furthermore move baryons to larger distances
away from the center of the galaxies (outside of the aperture), but
this has to be tested further by looking at the resolved \h2
properties of galaxies with IllustrisTNG. {\gp \citet{Stevens2018} find a similar drop at $z=0$ in the total cold
gas mass (\hi plus \h2) of IllustrisTNG galaxies at similar stellar
masses and also argue that an AGN feedback may be responsible for this.}

Putting the predicted \h2 mass in contrast to the ASPECS sensitivity
limit gives an idea of which galaxies might be missed by
ASPECS. At $z=0.29$ the ASPECS sensitivity limit is below the median
of the entire population of galaxies with stellar masses larger
than $10^{10}\,{\rm M}_\odot$ for the SC SAM and IllustrisTNG when adopting the
`Grav' aperture. When adopting the `3.5arcsec' aperture the situation
changes, and only the most \h2 massive galaxies are picked up by the
ASPECS survey (well above the median). The same conclusions are
roughly true at $z=1.43$. At $z=2.61$ the ASPECS sensitivity limit is
below the median of the galaxy population as predicted by the SAM for
galaxies with $M_* > 10^{11}\,\rm{M}_\odot$. The
ASPECS survey is sensitive to the galaxies with the
largest \h2 masses with stellar
masses in the range $10^{10}\,\rm{M}_\odot < M_* <
10^{11}\,\rm{M}_\odot$. Galaxies with lower stellar masses are
excluded by the ASPECS sensitivity limit, according to the predictions
by the SC SAM. The ASPECS sensitivity limit at $z=2.61$ is always above
the median predictions from IllustrisTNG, independent of the 
aperture. At $z=3.8$ the ASPECS sensitivity limit is always above the
median predictions by the models (both the SC SAM and
IllustrisTNG). According to the models, ASPECS is only sensitive to
galaxies with stellar masses $\sim10^{11}\,\rm{M}_\odot$ with the most
massive \h2 reservoirs (see Section \ref{sec:discussion} for a more in depth discussion on this).

\subsubsection{Mocked results}
In Figure \ref{fig:scale_evol_SAMvsTNG} we again present the \h2 mass of galaxies
as a function of their stellar mass at $z=0$ and at the median redshifts
of ASPECS predicted from IllustrisTNG and the SC SAM. Differently from the previous Figure, we now take into account the selection functions that characterize the observational datasets we compare to. In particular, in this Figure, the predictions are compared to observed \h2 masses of galaxies from \citet{Saintonge2017} at $z=0$, and to the detections from the
ASPECS surveys (all detections with a signal-to-noise ratio higher than 6.4). as well as a compilation presented in
\citet{Tacconi2018} as a part of the PHIBBS (IRAM Plateau de Bure
HIgh-z Blue Sequence Survey) survey at higher
redshifts. At $z=0$ a selection criterion of $\log{\rm{SFR}} >
  \log{\rm{SFR}_{\rm MS}}(\rm{M}_*) - 0.4$ is applied to both the
  observed and modeled galaxies, where $\log{\rm{SFR}_{\rm
      MS}}(\rm{M}_*) $ marks the SFR of galaxies on the main sequence
  of star-formation at $z=0$ following the definition of
  \citet{Speagle2014}. At higher redshifts we only adopt the ASPECS
  CO sensitivity based selection criterion. ASPECS is sensitive to
  sources with an \h2 mass of $\sim 10^9\rm\rm{M}_\odot$ at $z=0.29$
  and $\sim 10^{10},\,2\times 10^{10},\,$and
  $3\times10^{10}\,\rm{M}_\odot$, at $z\approx 1.43,\,2.61$ and $3.8$,
  respectively (see \citealt{Boogaard2019}, \citealt{Decarli2019}, and
  \citealt{Gonzalez2019}, for more details).\footnote{Like before, we adopt the same CO excitation conditions and CO--to--\h2
  conversion factor as ASPECS, $\alpha_{\rm CO} =
3.6\,\rm{M}_\odot/(\rm{K}\,\rm{km/s}\,\rm{pc}^{2})$, and assume a CO
line-width of 200 km/s. {\gp Note that one of the ASPECS sources in
  Figure \ref{fig:scale_evol_SAMvsTNG} has an \h2 mass below the
  dotted line representing the ASPECS selection function. This 
  galaxy has a CO line width narrower than 200 km/s. Accounting for
  variations in the CO line width heavily complicates the selection
  function that has to be applied to the IllustrisTNG and SC SAM
  galaxies. We have thus chosen to limit ourselves to a typical value
  for main-sequence galaxies of 200 km/s.}} {\gp PHIBBS selected
galaxies based on a lower-limit in stellar mass and SFR. The galaxies
in PHIBBS that have most massive \h2 reservoir also meet the ASPECS
criterium.}

At $z=0$ the predictions by the IllustrisTNG model are in general in
good agreement with the observations {\gp (\citealt{Diemer2019} presents a more detailed comparison of the \h2 mass properties of
galaxies at $z=0$ between model predictions and observations,
accounting for beam/aperture effects and different selection functions)}. The typical spread in the relation between \h2 mass and
stellar mass is smaller for the model galaxies than the observed
galaxies  (it is worthwhile to note that the sample size of
the observed galaxies is significantly smaller). At higher redshifts, on the other hand, a large fraction of the galaxies detected by ASPECS at $z \geq 1.43$ are not predicted by
either IllustrisTNG  (independent of the adopted
aperture) or the SC SAM, i.e., the observed galaxies lie outside of the
two-sigma scatter derived from the models. Similarly, a large fraction
of the galaxies that are part of the PHIBBS data compilation also lie
outside the {\gp two-sigma scatter on} the predictions by the different
models (also at $z\sim 3.8$). This suggests that the models predict \h2 reservoirs as a
function of stellar mass that are not massive enough at $z \sim
1-3$. 

Note that the median trends predicted from IllustrisTNG and the SC SAM at $z=0$ are essentially identical at low stellar masses, $\lesssim 10^{11}\,\rm{M}_\odot$. However, they diverge at larger stellar masses. The \h2 masses predicted from IllustrisTNG at $z=0$ are a factor $\sim2$ higher than the SC SAM's ones above $10^{11}\,\rm{M}_\odot$, the precise estimate depending on the adopted aperture. At $z\sim0.29$ the \h2 scaling relations predicted by the models when accounting for the ASPECS sensitivity limits begin to differ for galaxies with stellar masses larger than $\sim 7 \times 10^{10}\,\rm{M}_\odot$. At higher redshifts, the SC
SAM and IllustrisTNG predict similar \h2 masses for galaxies with
stellar masses less than $10^{11}\,\rm{M}_\odot$ {\gp (an artefact of
  the imposed selection limit)}, while at larger stellar
masses the SAM predicts slightly more massive \h2 reservoirs at fixed stellar mass. Overall, the predictions of the SC SAM and IllustrisTNG are surprisingly similar, considering the large number of differences in the underlying modeling approach.

\begin{figure*}
\includegraphics[width = \hsize]{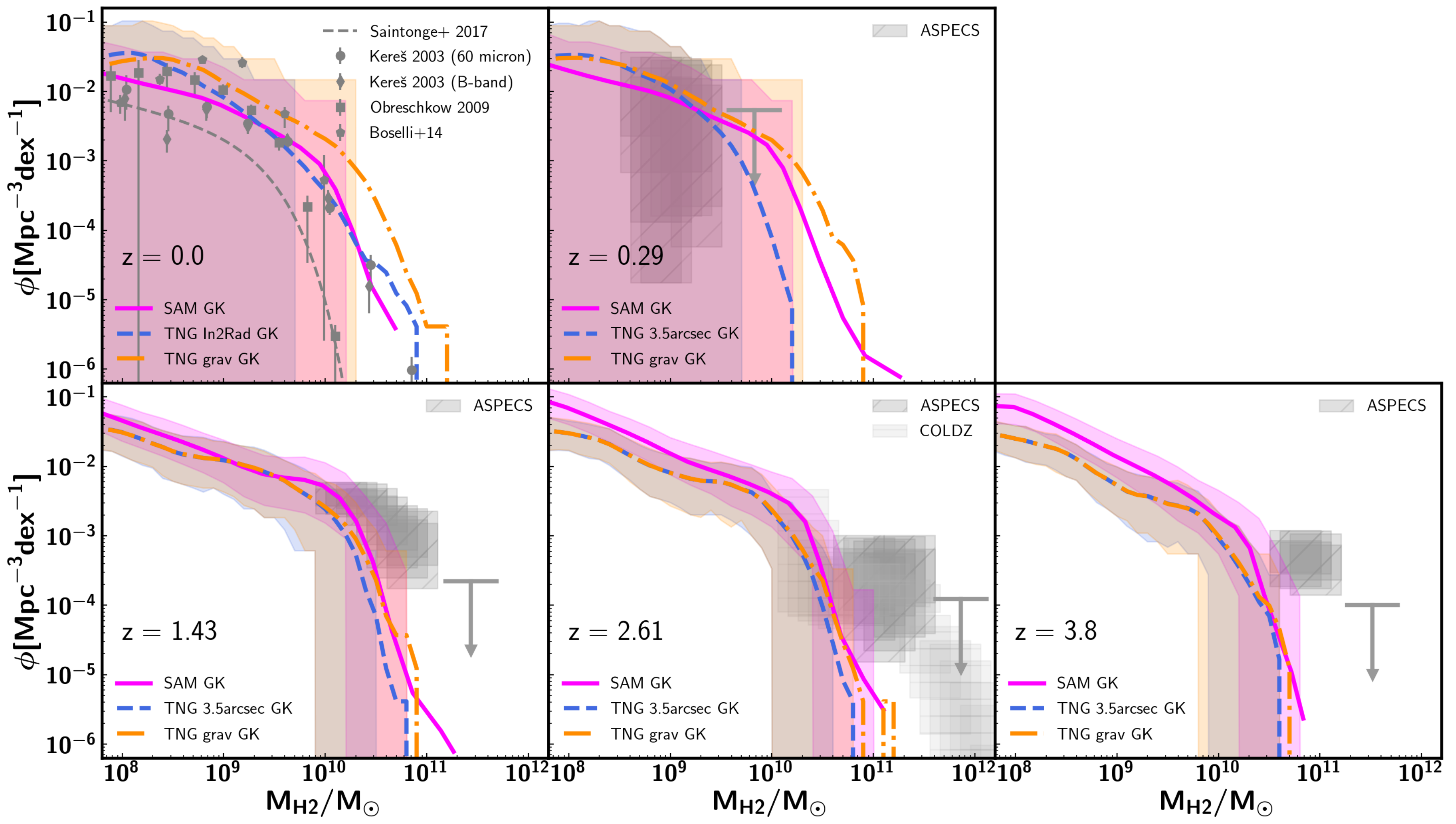}
\caption{The predicted and observed \h2 mass function of galaxies assuming $\alpha_{\rm CO} =
3.6\,\rm{M}_\odot/(\rm{K}\,\rm{km/s}\,\rm{pc}^{2})$ at $z=0$ and the redshifts probed by
  ASPECS. Model predictions are shown for the SC SAM (solid pink) and
  IllustrisTNG (`3.5arcsec' aperture: dashed blue; `Grav' aperture: dashed-dotted orange), both models adopting the GK \h2
  partitioning recipe. In this Figure the thick lines mark the mass function based on the entire
  simulated box ($\sim 100$ cMpc on a side for IllustrisTNG, $\sim 142$ cMpc on a side
  for the SC SAM). The colored shaded regions mark the {\gp two-sigma scatter}
  when calculating the \h2 mass function in 1000 randomly selected
  cones that capture a volume corresponding to the volume probed
  by ASPECS at the given redshifts (Table \ref{table_volumes}). At $z=0$ the model predictions are
  compared to observations from \citet{Keres2003},
  \citet{Obreschkow2009}, \citet{Boselli2014}, and \citet{Saintonge2017}. At higher redshifts
  the model predictions are
  compared to observations from the ASPECS and COLDZ
  \citep{Riechers2018}
  surveys.
\label{fig:massfunc_evol_SAMvsTNG}}
\end{figure*}

\subsection{The evolution of the \h2 mass function}
We show the \h2 mass function of galaxies as predicted from IllustrisTNG
 and the SC SAM for the GK \h2 partitioning recipe in Figure
 \ref{fig:massfunc_evol_SAMvsTNG} (the \h2 mass functions predicted
 using the other \h2 partitioning recipes are presented in Appendix
 \ref{sec:partitioning_recipes}, where we show that they are very similar). The \h2 mass functions are shown
at $z=0$ and at the median redshifts probed by ASPECS. The theoretical 
mass functions are derived by accounting for all the galaxies in the
full simulation box ($\sim 100$ cMpc, solid line). The shaded regions
mark the spread in the mass function when calculating it in smaller
boxes representing the ASPECS volume, which is further discussed in
Section \ref{sec:massfunc_variance}. The mass functions at $z=0$ are
compared to observations taken from from
\citet{Keres2003}, \citet{Obreschkow2009}, 
\citet{Boselli2014}, and \citet[assuming a CO--to--\h2 conversion
factor of $\alpha_{\rm CO} =3.6\,\rm{M}_\odot/(\rm{K}\,\rm{km/s}\,\rm{pc}^{2})$]{Saintonge2017}. The \citet{Obreschkow2009} and \citet{Keres2003}
mass functions are based on the same dataset, only \citet{Obreschkow2009} assumes a
  variable CO--to--\h2 conversion factor as a function of metallicity
  (unlike ASPECS) instead of a fixed CO--to--\h2
  conversion factor. At higher redshifts we compare the model
  predictions to the results from ASPECS, as well as the
  results from the COLDZ survey at $z\sim2.6$ \citep{Pavesi2018,Riechers2018}. 

The \h2 mass function at $z=0$ predicted by the SC SAM is in good agreement with the observations \citep{Keres2003,Boselli2014,Saintonge2017}.\footnote{The differences between the observational mass functions are driven by field-to-field variance, as these surveys target a relatively small area on the sky or sample, sometimes located in known overdensities.} The mass function as predicted from IllustrisTNG
when adopting the `In2Rad' aperture (similar to the observed aperture) is also in rough agreement with the
observations. When adopting the `Grav' aperture the number densities
of the most massive \h2 reservoir are instead too high.  This
difference highlights the importance of properly matching the aperture
over which measurements are taken, especially at low redshifts and at
the high mass end. \citet{Diemer2019}  presents a robust comparison between model predictions from IllustrisTNG and observations at $z=0$, better accounting for the beam size of the various observations at $z=0$ than is done in this work.

Both the SC SAM and IllustrisTNG reproduce the  the observed \h2 mass function by
ASPECS at $z\sim 0.29$ (independent of the aperture). These are at masses below the knee of the mass function. Indeed, the volume probed by
ASPECS at $z\sim 0.29$ is rather small, which explains the lack of
galaxies detected with \h2 masses larger than a few times
$10^9\,\rm{M}_\odot$.  
For the most massive \h2
reservoirs at $z=0.29$, on the other hand, the two models (and the
choice of different apertures) return significantly different results:
at fixed number density, the corresponding \h2 mass differs {\gp  by 
a factor of five} between the two IllustrisTNG
apertures, with the SC SAM in between.

At $z>1$ the predictions for the \h2 number densities by the different models and their respective
apertures are very close to each other. On average the SC SAM predicts
number densities that are $\sim 0.2$ dex higher. At $z=1.43$ the models only just reproduce the observed \h2 mass function around
masses of $10^{10}\,\rm{M}_\odot$, but predict too few galaxies with
\h2 masses larger than $3\times 10^{10}\,\rm{M}_\odot$. The predicted \h2 number densities at $z=2.61$ are in good agreement with COLDZ and ASPECS in the mass range $10^{10} \rm{M}_\odot \leq
M_{\rm H2} \leq 6\times 10^{10} \rm{M}_\odot $. The models do not
reproduce ASPECS at higher masses and at higher redshifts, predicting number densities that are too low. We will further quantify how well the models reproduce
the observed \h2 mass function when taking the surface area into account in the next subsection.

\begin{table}
\center
 \caption{The volume (in comoving Mpc) probed by ASPECS in different
   redshift ranges, {\gp after correcting for the primary beam sensitivty} \citep[see][]{Decarli2019}. \label{table_volumes}}
 \begin{tabular}{cc}
\hline
Redshift range & Volume (cMpc$^3$)\\
\hline
$0.003 \leq z \leq 0.369$ & 338\\
$1.006 \leq z \leq 1.738$ & 8198\\
$2.008 \leq z \leq 3.107$ & 14931\\
$3.011 \leq z \leq 4.475$ & 18242\\
\hline
\hline
 \end{tabular}
 \end{table}

\subsubsection{Field-to-field variance effects on the \h2 mass function}
\label{sec:massfunc_variance}
Since ASPECS only surveys a small area on the sky, field-to-field variance may bias
the observed number densities of galaxies towards lower or higher
values. In Figure \ref{fig:massfunc_evol_SAMvsTNG} the thick lines represent the \h2
mass function that is derived when calculating the \h2 mass function
based on the entire simulated volume ($\sim$ 100 cMpc for TNG100).  The shaded areas around the thick lines
in Figure \ref{fig:massfunc_evol_SAMvsTNG} quantify the effects of cosmic
variance on the \h2 mass function. The shaded regions mark the
{\gp two-sigma scatter}  when calculating the \h2 mass function in {\gp 1000 randomly selected
  cones through the simulated volume} that capture a volume corresponding to the actual volume probed
  by ASPECS at the given redshifts (Table
  \ref{table_volumes}).\footnote{\gp Note that these correspond to cones through a model snapshot and not an continuous lightcone.}

At $z=0.29$ the small area probed
by ASPECS can lead to large differences in the observed \h2 mass
function. This ranges from number densities less than
$10^{-6}\,\rm{Mpc}^{-3}\,\rm{dex}^{-1}$ at the {\gp lower end of the
  two-sigma scatter} to a few
times $10^{-2} \,\rm{Mpc}^{-3}\,\rm{dex}^{-1}$ at the {\gp upper end
  of the two-sigma scatter} at any \h2 mass. The galaxies with the largest predicted \h2 reservoirs
at $z=0.29$ ($M_{\rm H2} > 10^{10}\,\rm{M}_\odot$) will typically be
missed by a survey like ASPECS ({\gp do not fall in between the two-sigma scatter}). This is
indeed reflected by the lack of constraints on the number density of
galaxies with \h2 masses more massive than $10^{10}\,\rm{M}_\odot$ by
ASPECS.

The volume probed by ASPECS at redshifts $z>1$ is significantly larger
(see Table \ref{table_volumes}), which indeed results in less scatter in the \h2
number densities of galaxies due to field-to-field variance. The two-sigma scatter in the
power-law component of the mass function is 0.2--0.3 dex for
IllustrisTNG and the SC SAM. The scatter quickly increases at \h2 masses
beyond the knee of the mass functions, ranging from number densities
less than $10^{-6}\,\rm{Mpc}^{-3}\,\rm{dex}^{-1}$ to number densities
a few times higher than inferred based on the entire simulated
boxes. The model galaxies that host the largest \h2 reservoirs in the full modeled boxes are typically not recovered when focusing on small volumes similar to the volume probed
by ASPECS.

We can make a fairer comparison between the predictions by the theoretical models
and ASPECS by accounting for the small volume probed by
ASPECS. Figure \ref{fig:massfunc_evol_SAMvsTNG} shows that at $z>1$
the observed number density of galaxies with $M_{\rm H2} >
10^{11}\,\rm{M}_\odot$ is outside of the {\gp two-sigma
  scatter} of the model
predictions by both IllustrisTNG (for both apertures) and the SC
SAM. The number densities of galaxies with lower
\h2 masses are within the {\gp two-sigma scatter} of both models.
Summarizing, both IllustrisTNG and the SC SAM do not
predict enough \h2 rich galaxies (with masses larger than
$10^{11}\,\rm{M}_\odot$) in the redshift range $1.4 \leq z
\leq 3.8$. This is in line with our findings in Section
\ref{sec:scale_evol} that both IllustrisTNG and the SC
SAM predict \h2
masses within this redshift range that are typically too low for their
stellar masses compared to the observations from ASPECS and PHIBBS.

\begin{figure*}
\centering
\includegraphics[width = 0.7\hsize]{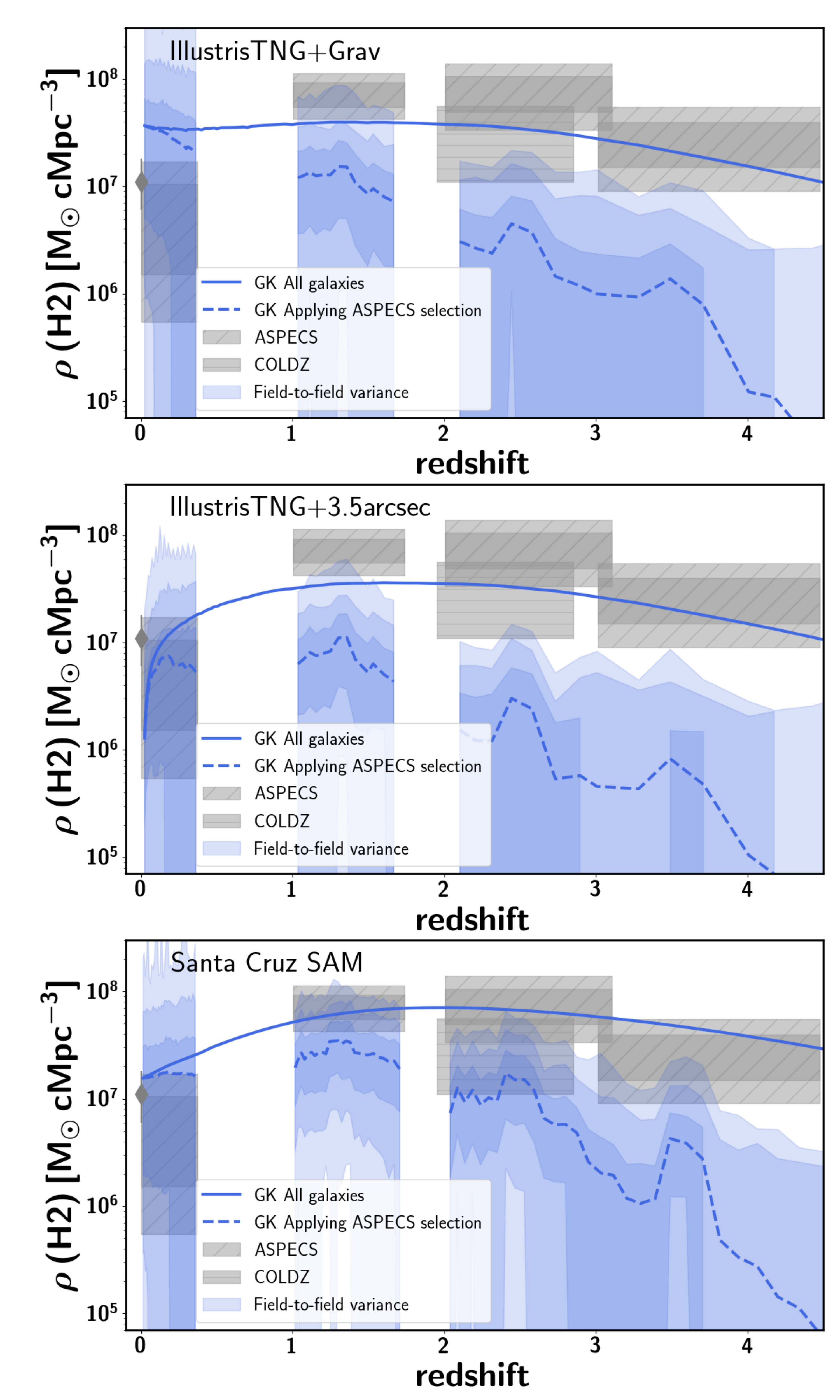}
\caption{The predicted and observed \h2 cosmic density assuming $\alpha_{\rm CO} =
3.6\,\rm{M}_\odot/(\rm{K}\,\rm{km/s}\,\rm{pc}^{2})$ as a function of
  redshift predicted from IllustrisTNG (`Grav' aperture, top; `3.5arcsec' aperture, center),  and the SC SAM (bottom),
  adopting the GK \h2 partitioning
  recipe. Solid lines correspond to the cosmic \h2 density based on all the galaxies in the entire simulated volume. Dashed lines correspond to the
  cosmic \h2 density when applying the ASPECS selection
  function. Shaded regions mark the 0th and 100th percentiles, {\gp
    two-sigma, and one-sigma scatter}  when calculating the \h2 cosmic density in 1000 randomly selected
  cones that capture a volume representing the ASPECS survey. Observations are from ASPECS (dark (light) grey mark the
  one (two) sigma uncertainty), COLDZ \citep{Riechers2018}, and
  from \citet{Saintonge2017} at $z=0$.  \label{fig:rho_evol_SAMvsTNG_variance}}
\end{figure*}

\subsection{The \h2 cosmic density}
\label{sec:cosmic_density}
We present the evolution of the cosmic density of \h2 within galaxies predicted by the SC SAM and IllustrisTNG when adopting the GK partitioning recipe in Figure \ref{fig:rho_evol_SAMvsTNG_variance} (predictions for the other partitioning recipes are presented in Appendix \ref{sec:partitioning_recipes}). The
solid lines correspond to the cosmic density derived based on all the
galaxies in the entire simulated volume. The dashed lines correspond
to a scenario where we only include galaxies with \h2 masses larger
than the detection limit of ASPECS. The shaded region marks the \h2 cosmic
density calculated in a box with a volume that corresponds to the volume probed by ASPECS at the appropriate redshift. This is further explained in Section \ref{sec:cosmic_density_variance}. The model predictions are compared to
$z=0$ observations taken from \citet{Keres2003} and \citet{Obreschkow2009},
as well as the observations from the ASPECS and COLDZ
\citep{Riechers2018} surveys at higher redshifts.

The \h2 cosmic density predicted from IllustrisTNG when adopting the
`Grav' aperture gradually increases till $z=1.5$ after which it stays
roughly constant till $z=0$. At $z\sim$1, accounting for the ASPECS
sensitivity limits can lead to a reduction in the \h2 cosmic density
of a factor of three. {\gp The reduction is already one order of magnitude at $z\sim2$ and further increases towards
higher redshifts.} The \h2 cosmic density predicted from IllustrisTNG when
adopting the `3.5arcsec' aperture increases till $z\sim 2$ and stays roughly constant till $z=1$. The \h2
cosmic density rapidly drops at $z<1$ by almost an order of
magnitude till $z=0$. The difference between the low-redshift evolution predicted
when adopting the `Grav' aperture versus the `3.5arcsec' aperture
(especially at $z<0.5$) indicates that the `3.5arcsec' aperture misses
a significant fraction of the \h2 associated with the galaxy. The
decrease in \h2 cosmic density when accounting for the ASPECS
sensitivity limits is similar for the `3.5arcsec' aperture as the
`Grav' aperture. The decrease is approximately a factor of three at
$z=1$, approximately an order of magnitude at $z=2$, and this increases
towards higher redshifts. 

The \h2 cosmic density as predicted by the SC SAM when including all galaxies increases till $z\sim2$, after which it gradually decreases by a factor of $\sim 4$ till $z=0$. Similar to IllustrisTNG, accounting for the ASPECS sensitivity limits results in a drop in the \h2 cosmic density of a factor of $\sim 3$ at $z=1$ and approximately an order of magnitude at $z>2$. On average, the SC SAM predicts \h2 cosmic
densities at $z>1$ that are 1.5--2 times higher than predicted from IllustrisTNG (note that the SC SAM also predicts higher number densities for \h2-rich galaxies at these redshifts).

The difference between the total cosmic density (i.e., including the contribution from all
galaxies in the simulated volume) and the \h2 cosmic density after applying
the ASPECS sensitivity limit highlights the importance of properly
accounting for selection effects when comparing model predictions to
observations. Too often, comparisons are only carried out at face value
ignoring these effects, creating a false impression. In this analysis we find that, when taking the ASPECS sensitivity limits into account, the
cosmic densities predicted by the models are well below the observations at $z>1$, independent of the adopted model, \h2 partition recipe, and
aperture. In the next subsection we will additionally take the effects of cosmic
variance into account, in order to better quantify the (dis)agreement between ASPECS
and the model predictions.

\subsubsection{Field-to-field variance effects on the \h2 cosmic density}
\label{sec:cosmic_density_variance}
To understand the effects of field-to-field variance on the results from the
ASPECS survey we also calculate the \h2 cosmic density in boxes representing the ASPECS volume. The shaded regions in Figure \ref{fig:rho_evol_SAMvsTNG_variance}  mark
the 0th and 100th percentiles, {\gp
    two-sigma, and one-sigma scatter}  when calculating the \h2
cosmic density in {\gp 1000 randomly selected cones through the simulated volume }that correspond to the
volume probed by ASPECS (also accounting
for the ASPECS sensitivity limit).\footnote{At some redshifts, for
  example $z>3.5$, the shaded area corresponding to the one-sigma
  scatter appears to be missing. At these redshifts the one-sigma area
  falls below the minimum \h2 density depicted in the figure and is therefore not shown.}

At $z<0.3$ field-to-field variance can lead to large variations
already ({\gp multiple orders of magnitude within the two-sigma scatter}) in the derived \h2 cosmic density of the
Universe, both for IllustrisTNG and the SC SAM. At higher redshifts
the volume probed by ASPECS is larger and indeed the scatter in the
\h2 cosmic density is smaller than at $z<0.3$.

The ASPECS observations at $z\sim1.43$ are reproduced by a
small fraction of the realizations predicted from IllustrisTNG
(independent of the aperture), corresponding to the area {\gp above the
two-sigma scatter} (i.e., only up  to 2.5\% of the realizations
drawn from IllustrisTNG reproduce the ASPECS observations). The
observations at $z\sim1.43$ are reproduced by a larger fraction of the
realizations drawn from the SC SAM, covering the area between {\gp the one-
and two-sigma scatter and above}.

At $2<z<3$ all the realizations drawn from IllustrisTNG {\gp
  (independent of the  chose aperture)} predict \h2 cosmic densities lower
than the ASPECS observations.  {\gp At $z>3$ only a small fraction of the
realizations reproduce the ASPECS observations when adopting the
`Grav' aperture, corresponding to the area between the two-sigma scatter and
100th percentile.} The SC SAM predicts 
slightly higher cosmic densities on average, and indeed the ASPECS observations at
$z>2$ {\gp fall within the two-sigma scatter of the realizations}. Both IllustrisTNG and the
SC SAM reproduce the observations taken from COLDZ in a subset of the realizations. 

It is important to realize that a model is
not necessarily ruled out if not all of the realizations agree with the
ASPECS results. The fraction of realizations that agrees with the
ASPECS results gives a feeling for the likelihood of a model being
realistic. If only a small fraction (or none) of the realizations reproduces
the ASPECS observations, this suggests that the model is very likely to be invalid (modulo the assumptions with regards to the interpretation of the observations). We will come back to this in the discussion.

\begin{figure*}
\includegraphics[width = \hsize]{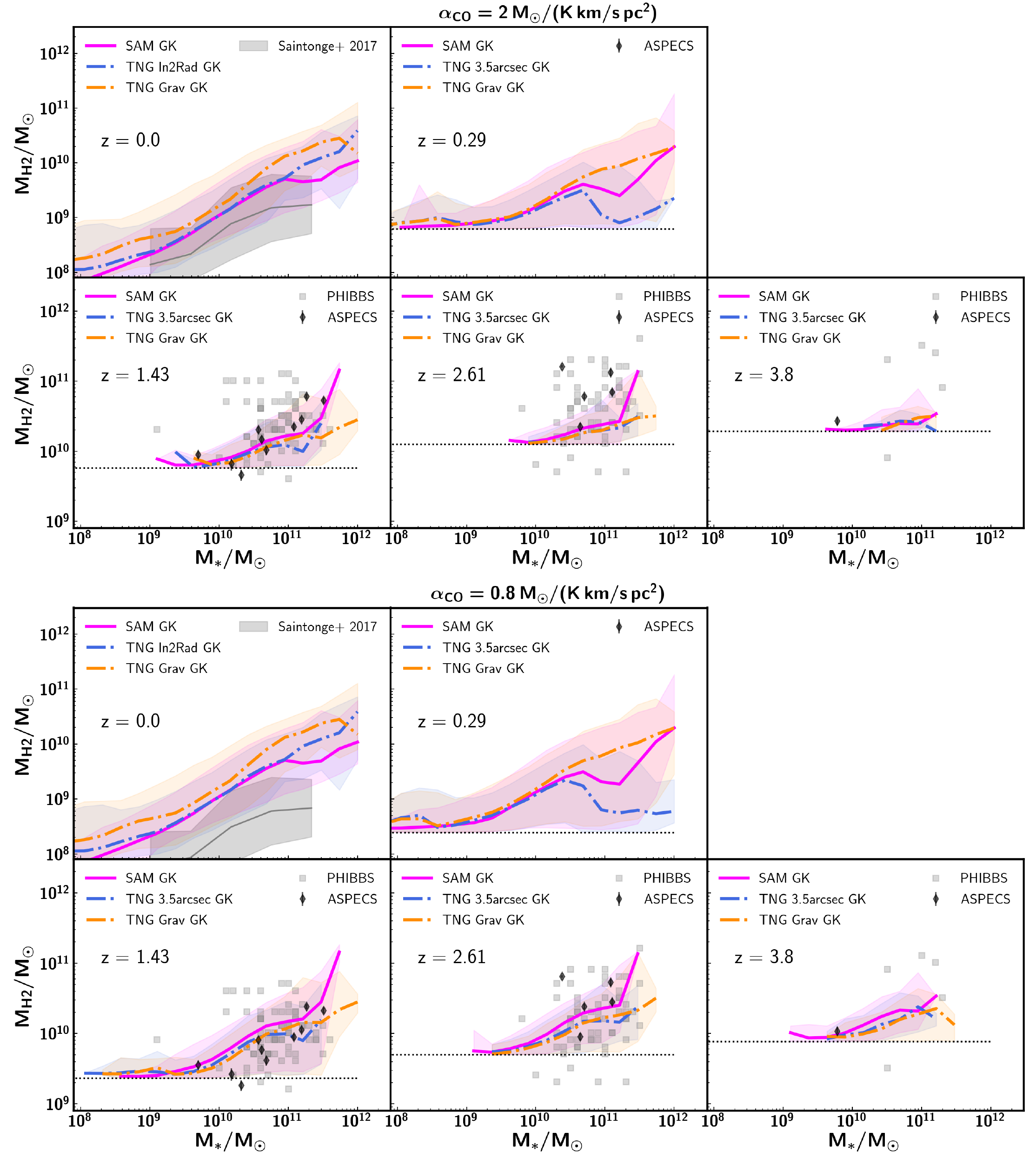}
\caption{The predicted and observed \h2 mass of galaxies at different redshifts as a function of their
  stellar mass. The top five and bottom five panels correspond to a scenario where we adopt a CO--to--\h2 conversion factor of $\alpha_{\rm CO} =
2.0\,\rm{M}_\odot/(\rm{K}\,\rm{km/s}\,\rm{pc}^{2})$ and $\alpha_{\rm CO} =
0.8\,\rm{M}_\odot/(\rm{K}\,\rm{km/s}\,\rm{pc}^{2})$ for the observations and the simulations (through the ASPECS selection function), respectively. This Figure is otherwise identical to Figure \ref{fig:scale_evol_SAMvsTNG}.
\label{fig:scale_evol_SAMvsTNG_alphaCO}}
\end{figure*}

\begin{figure*}
\includegraphics[width = \hsize]{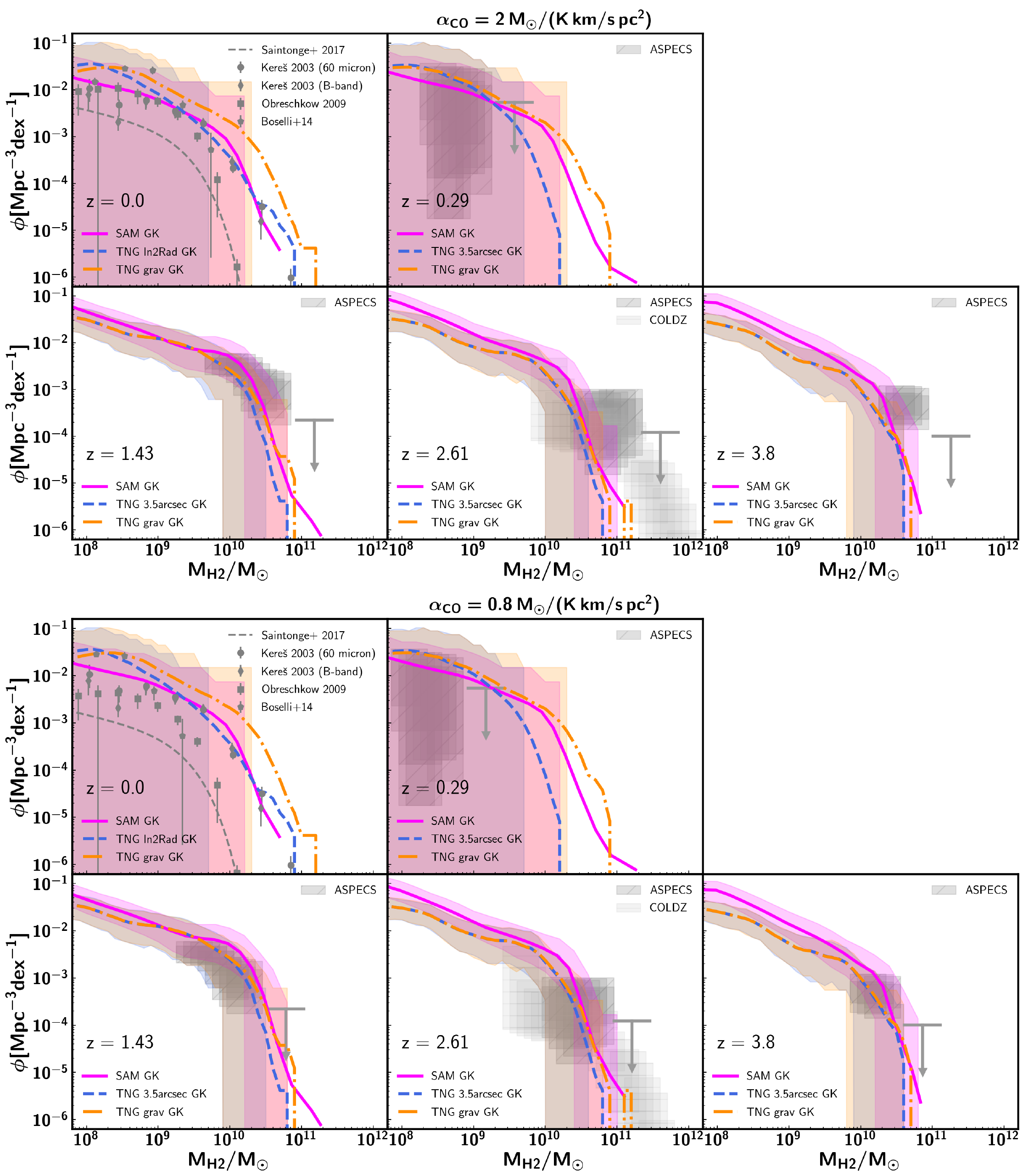}
\caption{The predicted and observed \h2 mass function of galaxies at $z=0$ and the redshifts probed by
  ASPECS as predicted from IllustrisTNG and the SC SAM. The top five and bottom five panels correspond to a scenario where we adopt a CO--to--\h2 conversion factor of $\alpha_{\rm CO} =
2.0\,\rm{M}_\odot/(\rm{K}\,\rm{km/s}\,\rm{pc}^{2})$ and $\alpha_{\rm CO} =
0.8\,\rm{M}_\odot/(\rm{K}\,\rm{km/s}\,\rm{pc}^{2})$ for the observations, respectively. The data comparison is identical to Figure
  \ref{fig:massfunc_evol_SAMvsTNG}. In this Figure the thick lines mark the mass function based on the entire
  simulated box (100 Mpc on a side for IllustrisTNG, 142 Mpc on a side
  for the SC SAM). The shaded regions mark the {\gp
    two-sigma scatter} when calculating the \h2 mass function in 1000 randomly selected
  cones that capture a volume corresponding to the volume probed
  by ASPECS at the given redshifts.
\label{fig:massfunc_alphaCO}}
\end{figure*}

\section{Discussion}
\label{sec:discussion}
\subsection{Not enough \h2 in galaxy simulations?}
One of the main results of this paper is that, when a CO--to--\h2 conversion factor $\alpha_{\rm CO} =
3.6\,\rm{M}_\odot/(\rm{K}\,\rm{km/s}\,\rm{pc}^{2})$ is assumed, both IllustrisTNG and
the SC SAM predict \h2 masses that are too low at a given
stellar mass for galaxies at $z>1$ (Figure \ref{fig:scale_evol_SAMvsTNG}), do not predict enough \h2-rich
galaxies (with \h2 masses larger than $3\times
    10^{10}\,\rm{M}_\odot$; Figure \ref{fig:massfunc_evol_SAMvsTNG}), and predict cosmic
densities that are only {\gp marginally compatible (SC SAM) or in tension (IllustrisTNG)} with the ASPECS
results after taking the ASPECS sensitivity limits into
 account (Figure \ref{fig:rho_evol_SAMvsTNG_variance}). There are multiple choices that have to be made (both from the theoretical and observational side) that will affect this conclusion. In the remainder of this sub-section we discuss the main ones.

\subsubsection{The strength of the UV radiation field impinging on molecular clouds}
One of the theoretical challenges when calculating the \h2 content of
galaxies is accounting for the impinging UV radiation
field. \citet{Diemer2018} explored multiple approaches, by increasing and decreasing the UV radiation field when calculating the
\h2 mass of cells in the IllustrisTNG simulation. The authors found
differences in the predicted \h2 masses within a factor of 3 for the
most extreme scenarios tested in their work (ranging from 1/10 to 10
times their fiducial UV radiation field, where a stronger UV radiation field results in lower \h2 masses), with differences away from
their fiducial model up to a factor of 1.5-2. Although  a
systematic decrease in the UV radiation field could help to reproduce the cosmic density of \h2,
it would go at the cost of reproducing the \h2 mass of galaxies and
their mass function at $z=0$. Furthermore, a factor of 1.5--2
higher \h2 masses would still not be enough to overcome the tension
between model predictions and observations at $z>2$. {\gp In the context of the SC SAM, SPT15 explored
two different approaches to calculate the strength of the UV radiation
field and found minimal changes in the predicted \h2 mass of galaxies with a $z=0$ halo mass larger than
$10^{11}\,\rm{M}_\odot$. }

\begin{figure*}
\centering
\includegraphics[width = \hsize]{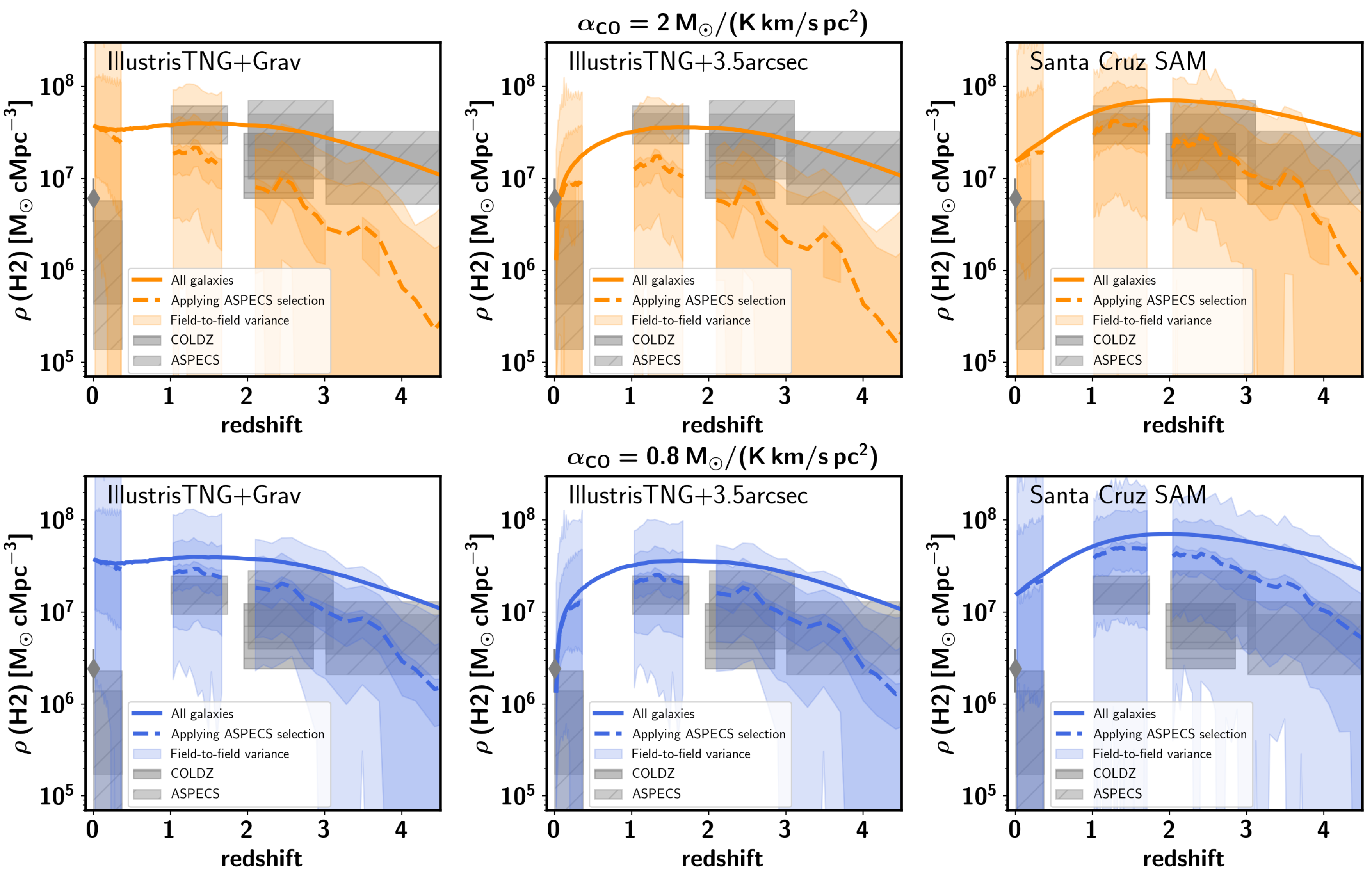}
\caption{Model predictions for the \h2 cosmic density as a function of
  redshift as predicted from IllustrisTNG adopting the `Grav' aperture
  (left), IllustrisTNG adopting the `3.5arcsec' aperture (middle),  and the SC SAM (right column),
  all of them adopting the GK \h2 partitioning
  recipe. The first and second rows correspond to a scenario where we adopt a CO--to--\h2 conversion factor of $\alpha_{\rm CO} =
2.0\,\rm{M}_\odot/(\rm{K}\,\rm{km/s}\,\rm{pc}^{2})$ and $\alpha_{\rm CO} =
0.8\,\rm{M}_\odot/(\rm{K}\,\rm{km/s}\,\rm{pc}^{2})$ for the observations, respectively. The solid lines corresponds to the cosmic \h2 density based on all the galaxies in the simulated volume,
  ignoring any selection function. The dashed lines correspond to the
  cosmic \h2 density in the entire box, applying the ASPECS selection
  function. The shaded regions mark the 0th and 100th percentiles, {\gp
    two-sigma, and one-sigma scatter} 
  when calculating the \h2 mass function in 1000 randomly selected
  cones that capture a volume corresponding to the volume probed
  by ASPECS at the given redshifts, also applying the
  ASPECS selection function. Model predictions are compared to the
  observational results from ASPECS (dark (light) gray mark the
  one (two) sigma uncertainty), as well as
  observations at $z=0$ from \citet{Keres2003} and
  \citet{Obreschkow2009}.  \label{fig:cosmic_evol_alphaCO}}
\end{figure*}

\subsubsection{The CO--to--\h2 conversion factor and CO excitation conditions}
One of the major observational uncertainties that could alleviate the
tension between model predictions and the ASPECS results is the
CO--to--\h2 conversion factor. The ASPECS survey adopts a conversion
factor of $\alpha_{\rm CO} =
3.6\,\rm{M}_\odot/(\rm{K}\,\rm{km/s}\,\rm{pc}^{2})$ for all CO
detections. We first explore what values for the CO--to--\h2
conversion factor would be necessary to bring the model predictions into
agreement with the observations. Changing the assumption for
$\alpha_{\rm CO}$ has two immediate consequences. Firstly, it changes
the value of the observed \h2 mass. Secondly, it changes the \h2 mass
limit below which galaxies are not detected (since observations have a
CO rather than an \h2 detection limit). Additionally it is important to better constrain the ratio between the CO J$=$1--0 and higher order rotational transitions of CO (J$=$2--1 to J$=$4--3 in the ASPECS survey). This ratio is currently assumed to be a fixed number, but has been shown to vary by a factor of a few from Milky Way type galaxies to ULIRGS.

We show the \h2 mass of galaxies as a function of stellar mass when varying the CO--to--\h2 conversion factor in Figure \ref{fig:scale_evol_SAMvsTNG_alphaCO}. The model predictions at $z=1.43$ are in significantly better agreement with the ASPECS detections when adopting $\alpha_{\rm CO} =
2.0\,\rm{M}_\odot/(\rm{K}\,\rm{km/s}\,\rm{pc}^{2})$ than the standard value of $\alpha_{\rm CO} =
3.6\,\rm{M}_\odot/(\rm{K}\,\rm{km/s}\,\rm{pc}^{2})$, although there
are still a significant number of galaxies detected as part of the
PHIBBS survey with \h2 masses outside of the {\gp
  two-sigma scatter} of the models. More than half of the ASPECS
detections at $z=2.61$ fall outside of the {\gp two-sigma scatter} of the model predictions when adopting $\alpha_{\rm CO} =
2.0\,\rm{M}_\odot/(\rm{K}\,\rm{km/s}\,\rm{pc}^{2})$. When assuming $\alpha_{\rm CO} =
0.8\,\rm{M}_\odot/(\rm{K}\,\rm{km/s}\,\rm{pc}^{2})$, the model predictions are in good agreement with the ASPECS detections at $z=1.43$ and $z=2.61$ (although there are still a number of PHIBBS detections not reproduced by the models). We do note that the better match at $z>1$ comes at the cost of predicting \h2 masses that are  too massive at $z=0$.

We present the observed and
predicted \h2 mass function of galaxies when assuming different values
for $\alpha_{\rm CO}$ in Figure \ref{fig:massfunc_alphaCO}. We find
that when adopting $\alpha_{\rm CO} =
2.0\,\rm{M}_\odot/(\rm{K}\,\rm{km/s}\,\rm{pc}^{2})$ the models
reproduce the observed ASPECS \h2 {\gp mass function of galaxies over
cosmic time (after accounting for cosmic variance,} Figure
\ref{fig:massfunc_alphaCO} top panels vs. Figure
\ref{fig:massfunc_evol_SAMvsTNG}). {\gp The number density of massive
  galaxies (larger than $10^{11}\,\rm{M}_\odot$)  detected as a part of
  the COLDZ survey are still not reproduced by the models (i.e., the observed number
densities are outside the {\gp two-sigma scatter} of the model
predictions)}. A CO--to--\h2 conversion factor of $\alpha_{\rm CO} =
0.8\,\rm{M}_\odot/(\rm{K}\,\rm{km/s}\,\rm{pc}^{2})$ brings the model
predictions for the \h2 mass functions from IllustrisTNG and the SC
SAM into excellent agreement with the results from ASPECS at $1
\lesssim z \lesssim 4$ (Figure \ref{fig:scale_evol_SAMvsTNG_alphaCO},
lower panels) and yields best agreement with the COLDZ results.

{\gp When adopting $\alpha_{\rm CO} =
2.0\,\rm{M}_\odot/(\rm{K}\,\rm{km/s}\,\rm{pc}^{2})$, both models return a larger
fraction of volume realizations that are consistent with the ASPECS and COLDZ \h2 cosmic densities at all redshifts
(Figure \ref{fig:cosmic_evol_alphaCO}, top panels).  The ASPECS
observations fall well within the two-sigma scatter of the predictions
by the SC SAM.  The observations fall in the area between the
two-sigma scatter and 100th percentile of the predictions by
IllustrisTNG when adopting an aperture corresponding to 3.5 arcsec. When adopting $\alpha_{\rm CO} =
0.8\,\rm{M}_\odot/(\rm{K}\,\rm{km/s}\,\rm{pc}^{2})$, the ASPECS
results match the predictions by both models (and both apertures for
IllustrisTNG). For this scenario, only the lower 32\% of all the
realizations predicted by the SC SAM matches the observations. Similar conclusions hold when comparing the model predictions to the COLDZ survey. We do note that reproducing the ASPECS results at $z>1$ by varying the CO--to--\h2 conversion factor for all galaxies comes at the cost of predicting \h2 masses for galaxies at $z=0$ that are too massive.}

Summarizing, the ASPECS survey would need to adopt a conversion factor of $\alpha_{\rm CO} \sim
0.8\,\rm{M}_\odot/(\rm{K}\,\rm{km/s}\,\rm{pc}^{2})$ for all observed
galaxies at $z>0$ for the models to better reproduce the observed \h2 mass function and the \h2 cosmic density.  The CO--to--\h2 conversion factor adopted by ASPECS of $\alpha_{\rm CO} =
 3.6\,\rm{M}_\odot/(\rm{K}\,\rm{km/s}\,\rm{pc}^{2})$ is motivated for
 main-sequence galaxies based on dynamical
 masses \citep{Daddi2010}, CO line spectral energy distribution (SED) fitting
 \citep{Daddi2015} and solar metallicity $z>1$ main-sequence galaxies
 \citep{Genzel2012}. Nevertheless, conversion factors of $\alpha_{\rm
   CO} \sim 2 \,\rm{M}_\odot/(\rm{K}\,\rm{km/s}\,\rm{pc}^{2})$ have
 been found for main-sequence galaxies at $z= 1-3$
 \citep[e.g.,][]{Genzel2012,Popping2017cSFG}, also justifying the use of
 this value. A ULIRG CO conversion factor of $\alpha_{\rm
   CO} \sim 0.8 \,\rm{M}_\odot/(\rm{K}\,\rm{km/s}\,\rm{pc}^{2})$ seems
 unrealistic for the entire sample, although it is not ruled out that,
 for example, the CO brightest sources in the ASPECS survey have a CO--to--\h2 conversion factor close to a ULIRG value. In reality, the CO--to--\h2 conversion factor will likely depend on a combination of ISM conditions and the gas-phase metallicity \citep{Narayanan2012, Renaud2018} and vary between galaxies. 
 
The COLDZ survey directly targets that CO J$=$1--0 emission line. Therefore no assumptions have to be made on the CO excitation conditions. The two models predicted in this work are in somewhat better agreement with COLDZ than ASPECS, although the models do not reproduce the \h2 massive galaxies found as a part of COLDZ either. The tension between the presented models and the observations can therefore not be fully accounted for by CO excitation conditions. 
 
What this ultimately demonstrates is that there appears to be tension between the ASPECS survey results and model predictions, but a
better quantification of this tension requires a better knowledge
of the CO--to--\h2 conversion factor and a comparison between theory and
observations by looking at CO directly. Zoom-simulations have
suggested that $\alpha_{\rm
   CO}$ varies as a function of metallicity and gas surface density
 \citep{Narayanan2012}. Such variations will have an influence on the
 slope of the \h2 mass--stellar mass relation and the \h2 mass
 function, possibly further reducing the presented tension between
 observations and simulation. A number of cosmological semi-analytic models of
 galaxy formation have been coupled to carbon chemistry and radiative transfer codes in order to provide direct predictions for the CO luminosity of individual galaxies in cosmological volumes \citep{Lagos2012,Popping2016,Popping2018}. This approach bypasses the use of a CO--to--\h2 conversion factor to convert the observed quantities into \h2 masses. In line with our conclusions on the \h2 mass function, these models fail to reproduce the number of CO-bright sources detected by the ASPECS survey \citep{Decarli2016,Decarli2019}.

\subsection{Field-to-field variance and selection effects for ASPECS}
Although ASPECS is providing a completely new view on the budget of
gas available for star formation in the Universe, the conclusions from
this survey are limited by the achievable survey design. The ASPECS survey only
probes an area of 4.6 arcmin$^2$ on the sky. Although the survey marks the deepest effort of this kind so far, it is by far not large enough to
overcome significant uncertainties due to field-to-field variance. Simulations are ideally
suited to address how big the uncertainty in the derived
conclusions is due to field-to-field variance. 

In Section \ref{sec:massfunc_variance} we showed that the \h2 mass number densities
derived for galaxies at $z>1$ when accounting for the volume of the ASPECS
survey typically vary within a factor of two
from the mass function derived for the entire simulated box. If we try
to translate this to ASPECS, the `real' \h2 mass function of the Universe might have number densities a factor of two lower/higher
than measured as part of ASPECS. This number actually
increases as a function of \h2 mass (because more massive galaxies are more strongly clustered), leading to possibly larger
discrepancies for galaxies with \h2 masses of the order
$10^{11}\,\rm{M}_\odot$. 

A similar conclusion holds for the cosmic density of molecular
hydrogen (see Section \ref{sec:cosmic_density_variance}). Independent of
the underlying model, the one-sigma scatter in the \h2 cosmic density at $z>1$
when applying the ASPECS sensitivity limits is typically within a
factor of three from the cosmic density derived over the entire simulated
volume (the two-sigma scatter is significantly larger). This number increases to a factor of 5 at the highest
redshifts probed by ASPECS. At face value this means that the real
cosmic density of \h2 may be up to a factor of three lower/higher than
suggested by the observations so far. It is good to keep in mind that given that the models do not perfectly match the observed \h2 masses, our field-to-field variance statements may be incorrect as well. For reference, when no selection on \h2 is applied to the models, the typical one-sigma field-to-field variance-driven uncertainty is approximately 50\% at $z>1$ for IllustrisTNG and the SC SAM.

It is hard to further quantify if the `real' \h2 cosmic density (modulo the ASPECS CO selection function) is
indeed a factor of 3--5 lower/higher than currently observed without
any additional knowledge of the UDF. Spectroscopic observations of the UDF have
suggested that the UDF is over-dense at $z\sim
0.67,\,0.73,\,1.1,\,\rm{and}\,1.61$
\citep{Vanzella2005}. \citet{Bouwens2007} find
that the UDF is slightly under-dense at redshifts 3--5 (up to a factor
of 1.5). Additional observations surveying a larger area on the sky at
different locations will be necessary to properly bracket the expected
variations in the \h2 cosmic density by field-to-field variance. Tests with
the two simulations discussed in this work have shown that an increase
in the covered area by an order of magnitude (ideally by looking at different regions on the sky) brings down the field-to-field variance
driven uncertainty in the \h2 cosmic density to a factor of two at the two-sigma level and 30\% at the one-sigma level.

In Section \ref{sec:cosmic_density} we showed that, at least according to the models discussed in this work, a significant fraction of the cosmic \h2
budget is missed by the ASPECS survey. At $z=1$ this is a factor of
three, whereas at $z>2$ this already corresponds to 90\% or even
more. Based on a study of the CO luminosity function, \citet{Decarli2019} estimate that the ASPECS survey accounts
for approximately 80 \% of the total CO luminosity emitted by
galaxies. This is in stark contrast with model predictions, but fits
with the idea that the models predict \h2 masses that are too low (and therefore less of the total \h2 density is picked up by ASPECS, or
different CO--to--\h2 conversion factors and/or excitation conditions need to be applied). On top of this, the low-mass slope of the \h2 mass function at $z>2$ as predicted by the models in this paper is steeper than the slope assumed in \citet[who adopt the same slope as \citealt{Saintonge2017}]{Decarli2019} for the CO luminosity function.

\subsection{A comparison to earlier works}
The finding that theoretical models do not predict enough
\h2 in galaxies when adopting $\alpha_{\rm CO} =
3.6\,\rm{M}_\odot/(\rm{K}\,\rm{km/s}\,\rm{pc}^{2})$ is not new. \citet{Popping2015SHAM,Popping2015candels}
reached the same conclusion for the SC SAM. The biggest difference to
the work presented in this paper is that the authors compared the SC
SAM predictions to inferred \h2 masses (and a sub-set of the PHIBBS galaxies also shown in this work), which come with their own
uncertainties based on the underlying model and can lead to false
conclusions. 

\citet{Decarli2016,Decarli2019} found a disagreement between the
observed and modeled CO luminosity functions (a proxy for the \h2 mass
function) at different redshifts,
comparing the ASPECS CO luminosity functions to predictions from
\citet{Lagos2012} and the SC SAM
\citep{Popping2016}. The authors found that the models do not predict enough CO bright galaxies at $1\leq z \leq 3$. In the current work
we presented a more robust comparison, taking field-to-field variance effects
into account to better quantify the disagreement in the number of \h2-massive (CO-bright) galaxies. Furthermore, the
uncertainty in the observed \h2 mass function and cosmic density used in the current work are tighter than in
\citet{Decarli2016}.

\citet{Decarli2016}  also presented a comparison between the \h2 cosmic
density as derived from the ASPECS pilot survey and predictions by three
semi-analytic models \citep{Obreschkow2009,Lagos2011,Popping2014}. \citet{Decarli2016} showed that these SAMs are able to reproduce the \h2
cosmic density up to $z=4$. The semi-analytic model presented in \citet{Xie2017} reproduces the \h2 cosmic densities from \citet{Decarli2016} from $z=0$--$4$. \citet{Lagos2018} predicts \h2 cosmic densities in agreement with the \citet{Decarli2016} results for galaxies at $z<1$ and $z>3$, but below the \citet{Decarli2016} results at redshifts $1\leq z \leq 3$. These
predictions by SAMs seem very encouraging,  but the comparisons were incorrect as the model predictions did not account for the ASPECS
sensitivity limits. We have shown in Section \ref{sec:cosmic_density}
that accounting for the ASPECS sensitivity limits can lead to a
reduction of a factor of up to 10 in the \h2 cosmic density (depending
on the considered model and redshift). 

\citet{Lagos2015} presented predictions
for the cosmic density of \h2 based on the EAGLE \citep{Schaye2015} simulations. The \h2 cosmic density
predicted by \citet{Lagos2015} is only barely in agreement with the
available results at that time from
\citet{Walter2014}. \citet{Lagos2015} do not directly account for the sensitivity limits of the \citet{Walter2014}
observations, but do present the \h2 cosmic density when only
considering galaxies with \h2 masses larger than $10^9\,\rm{M}_\odot$.
This reduced their cosmic \h2 density by approximately 0.2 dex at
$z<3$, and even more at higher redshifts, up to an order of
magnitude at $z=4$. Applying the ASPECS sensitivity limits to the \citet{Lagos2015} model would further lower the predicted \h2 cosmic density. \citet{Lagos2015} furthermore does not reproduce
the observed \h2 mass function from \citet{Walter2014} and
predicts molecular hydrogen fractions lower than suggested by
\citet{Tacconi2013} and \citet[although using different selection
criteria than the samples presented in these works]{Saintonge2013}. 

\citet{Dave2017} provide predictions for the \h2 cosmic density
based on the MUFASA simulation
\citep{Dave2016}. \citet{Dave2017} find a peak in their \h2
cosmic density at $z\sim 3$ after which the cosmic density
decreases by a factor of three. The predicted densities based on all the galaxies in their simulated volume (not accounting for any sensitivity limit) are also significantly lower than the ASPECS results.

Putting all of these works together we can draw multiple
conclusions. First of all, the added value of this paper is that it
presents the most detailed comparison between model predictions and
observations on this topic to date. On the one hand because it is based on the
deepest CO survey to date, on the other hand because
it accounts for sensitivity limits, field-to-field variance effects, and
brackets systematic theoretical uncertainties (two different galaxy
formation model approaches and different approaches for the
partitioning of \h2).  Second,  a large number of galaxy formation
models based on different methods (hydrodynamic and semi-analytic models) predict \h2 cosmic densities, \h2 masses, and \h2
mass functions that are too low compared to the observations. A better quantification of the latter will require constraining the CO--to--\h2 conversion factor in galaxies. Alternatively, a more precise comparison will require direct predictions of the CO luminosity of galaxies by galaxy formation models.

\subsection{Putting the lack of \h2 in a broader picture}
In this sub-section we aim to put the apparent lack of \h2 (the fuel for SF) in a broader picture by qualitatively discussing how predictions for the SFR of galaxies by different models agree with observations. A fair comparison would account for the different SF tracers used in the observations (and the average time scales over which they trace SF) as well as survey depth and survey area. Such a comparison should simultaneously also take into account the differences between the galaxies that ASPECS is sensitive to versus surveys focusing on other galaxy properties. Such a comparison should furthermore take into account that the spatial apertures and the time-scales a SFR tracer is sensitive to (e.g, up to $\sim$ 0.1 Gyr for UV based tracers) may be different from the  spatial extent and instantaneous nature of a CO detection. Such a detailed comparison is beyond the scope of this work,
we therefore limit ourselves to a brief qualitative discussion of SFR predictions in
the literature where these effects were not taken into
account. For example, many theoretical SFRs listed in the literature often represent the instantaneous SFR of gas taken directly from simulations.

The notion that galaxy formation models predict galaxies with \h2
masses that are too low at $z>1$ for their stellar mass is consistent with a broader picture of challenges
for galaxy formation and evolution theory. For example, \citet{SomervilleDave2015} compared the predicted SFR of galaxies as a function of their stellar mass at $z>1$ for a wide range of galaxy
formation models (including SAMs and hydrodynamical models) to
observed SFRs. All the models considered in this compilation predict
SFRs a factor of 2--3 lower than suggested by the majority of  observations at
$z=1-3$, while exhibiting better agreement at lower redshifts. The
same conclusion holds for IllustrisTNG (see detailed discussions in \citealt{Donnari2018}). If the \h2 masses of modelled galaxies are too low for their
stellar mass, it is not surprising that the SFRs of these galaxies are
also too low when a molecular hydrogen based SF recipe is
adopted. This is not necessarily true for models that adopt a total-cold gas
based SFR recipe. However, the lack of \h2 suggests that there is
either not enough gas or this gas is not dense enough to become
molecular. A logical consequence is that this also leads to SFRs that
are too low.

Since the \h2 cosmic densities predicted from IllustrisTNG and the SC
SAM {\gp when assuming $\alpha_{\rm CO} =
3.6\,\rm{M}_\odot/(\rm{K}\,\rm{km/s}\,\rm{pc}^{2})$ are in tension} with the ASPECS observations, one would naively expect that
the cosmic SFR density (cSFR, the SFR density of the universe)
predicted from the models discussed in this work
is also too low compared to the observations (if the SFR represent an
instantaneous conversion from \h2 (gas) into stars). In
  \citet{Pillepich2018}, an ``at face-value'' comparison between the
  cosmic SFR  predicted from IllustrisTNG and the data compilation presented in \citet{Behroozi2013} reveals a factor $\sim$2 discrepancy at redshifts $1 \leq z \leq 3$ (note however that Pillepich et al. did not attempt to apply any observational mock post processing to simulated galaxies or take other survey specifics into account). SPT15 reproduces
the data compilation in \citet{Behroozi2013} well in the redshift
range $1\leq z \leq 3$. \citet{Yung2018} compares the cSFR
predicted from the SC SAM to higher redshift observations and finds good
agreement with the observational compilation (\citealt{Yung2018} does
include a UV luminosity sensitivity limit when calculating the cSFR to
allow for a fair comparison to the observed cSFR). It is possibly
surprising that the marginal agreement in the \h2 cosmic
density predicted by the SC SAM does not result in a cSFR that is too low at $1 \leq z \leq 3$, especially since
the SFR of galaxies as a function of stellar mass is not
reproduced. We again emphasize that in this redshift range
observational selections were not taken into account in the
comparison of the cSFR. A closer look at the results presented in SPT15 shows that
the SC SAM predicts too many galaxies with a stellar mass below the
knee of the stellar mass function at $1 \leq z \leq 3$. The
contribution of these galaxies to the total cSFR can (partially) explain the
agreement between the predicted and observed cSFR, despite the
disagreement in the \h2 cosmic density. This immediately demonstrates
that a fair comparison taking selection functions and survey design
into account is always important and necessary. It also demonstrates why integrated cosmic mass density is difficult to interpret -- small changes in the abundance of low-mass objects can make a significant difference.

\citet{Dave2016} find that MUFASA predicts a total cSFR (not
applying any selection functions and adopting the instantaneous SFR from the simulation) that is lower than the observed cSFR
at redshifts $z=$1--3. \citet{Furlong2015} finds that the total cSFR
predicted by EAGLE (again not accounting for selection effects and adopting the instantaneous SFR from the simulation) is
systematically 0.2 dex below  the observed cSFR at $z<3$. This
suggests that also for these simulations the disagreement between
observed and modeled cSFR can (at least partially) be explained by a
lack of \h2 (star-forming) gas. 

{\gp   It is useful to keep in mind that even though the predicted star-forming main sequence
and cSFR by different models appears to be in tension with
observations, the same models find much better agreements with observational constraints on
the galaxy stellar mass functions and the stellar mass density at the
corresponding redshifts and masses (see e.g. \citealt{Somerville2015}
for the SC SAM, \citealt{Furlong2015} for EAGLE, and \citealt{Donnari2018} for IllustrisTNG, and discussions therein). This surprising
mismatch could hint to issues in the comparisons
(e.g. selection effects and different galaxy masses contributing to
the different observables, and differently so at different cosmic
times), problems of self-consistency in the observational data
(\citealt{Madau2014} find that the intergral of the observed cSFR and
the stellar mass density disagree by about a factor of two with each
other, but see \citealt{Driver}), issues
in the way star formation is modeled \citep[e.g.,][]{Leja2018} and proceeds within simulated galaxies, or a combination of all.}

Isolating the underlying physical mechanism that is responsible
  for the lack of massive \h2 reservoirs compared to the ASPECS survey
  is not straightforward. Within galaxy formation models, different physical processes acting on the
  baryons work in concert to shape galaxies. Changing the
  recipe for one of these processes with the aim of better reproducing
  a specific feature of galaxies can result in a mismatch for
  some other feature of galaxies. On top of that, different models
  often have different prescriptions for the physical processes acting
  on baryons in galaxies (even when they are similar in nature, the
  specifics may differ). 
  
  A number of models have attempted to alter their recipes for stellar
feedback and the re-accretion of gas to better reproduce the SFR of
galaxies at a given stellar mass over cosmic time
\citep{Henriques2015,White2015,Hirschmann2016}. These efforts have demonstrated that
changes to the re-accretion of ejected matter \citep{Henriques2015,White2015} or a
strongly decreasing efficiency of stellar feedback \citep[see also the
implementation in \citealt{Pillepich2018}]{Hirschmann2016} are promising,
  but not sufficient to solve the SFR
  discrepancy. We argue that besides the stellar mass and SFR of galaxies, a successful model will additionally have to address the lack of \h2 
  discussed in this paper. \citet{Hirschmann2016} and
  \citet{White2015} indeed showed that a delayed re-accretion and
  decreasing efficiency of stellar feedback with time lead to better agreement with the inferred \h2 masses of $z>1$
  galaxies available at that time. This makes these changes promising,
  but more systematic theoretical exploration is needed. Additional venues
  to  (simultaneously) explore include changes in the star-formation recipes to allow
  for a wider range in star-formation efficiencies.

\section{Summary \& Conclusions}
\label{sec:conclusions}
In this paper we have presented predictions from IllustrisTNG (specifically the TNG100 volume) and the SC
SAM for the \h2 content of galaxies, the \h2 mass function, and the
\h2 cosmic density over cosmic time. These predictions were compared
to results from ASPECS and COLDZ, specifically focusing on two
issues; 1) how well do the models reproduce the results from the ASPECS
survey; 2) how do field-to-field variance and the ASPECS sensitivity limits
affect the results of ASPECS? We summarize our main results
below:
\begin{itemize}
\item When adopting the canonical CO--to--\h2 conversion factor of $\alpha_{\rm CO} =
3.6\,\rm{M}_\odot/(\rm{K}\,\rm{km/s}\,\rm{pc}^{2})$, the typical \h2 masses of galaxies at $z>1$ as a function of their stellar
  mass predicted from IllustrisTNG and the SC SAM are lower than the
  observations by a factor of 2--3. A significant number of galaxies detected
  as a part of ASPECS fall outside of the two-sigma scatter
  of these models.
\item IllustrisTNG and the SC SAM do not reproduce the number of
  \h2-rich galaxies observed by ASPECS at $z>1$ (not enough galaxies with \h2 masses larger than $\sim 3 \times 10^{10}\,\rm{M}_\odot$). 
\item The predictions by the SC SAM and IllustrisTNG for the \h2 cosmic density only just agree with the ASPECS results after taking field-to-field variance effects into account. This suggests that the predicted cosmic densities are too low.

\item After applying the ASPECS sensitivity limit, the \h2 cosmic density is a factor of three (an order of magnitude) lower at $z=1$ ($z>2$) than the \h2 cosmic density obtained when accounting for all simulated galaxies (independent of the model).

\item Adopting a global CO--to--\h2 conversion factor in the range $\alpha_{\rm CO} =
2.0 - 0.8\,\rm{M}_\odot/(\rm{K}\,\rm{km/s}\,\rm{pc}^{2})$ would
alleviate much of the tension between model predictions by
IllustrisTNG and the SC SAM and the ASPECS results {\gp(although a uniform
value of $\alpha_{\rm CO}
=0.8~\rm{M}_\odot/(\rm{K}\,\rm{km/s}\,\rm{pc}^{2})$ appears unlikely)}.  Likewise, a global change in the CO excitation conditions could alleviate some of the tension between models predictions and observations.

\item Because ASPECS has a small survey area, field-to-field variance can lead to variations of typically up to a factor of three in the derived number densities for the \h2 mass function and
  cosmic density. It is thus crucial that besides sensitivity limits, field-to-field variance
  effects are also taken into account when comparing model predictions to observations.  According to the outcome of the models discussed in this work, increasing the survey area by an order of magnitude would reduce the typical two-sigma uncertainty in the \h2 cosmic density due to field-to-field variance to a factor of 2 (one-sigma uncertainty is $\sim30$ per cent).
  
  \item The systematic uncertainty between different \h2 partitioning
  recipes for predictions of the \h2 mass of galaxies, the \h2 mass
  function, and the \h2 cosmic density of the Universe is minimal.
\end{itemize}

The result that IllustrisTNG and the SC SAM do not predict enough
\h2-rich galaxies at $z>1$ when adopting $\alpha_{\rm CO} =
3.6\,\rm{M}_\odot/(\rm{K}\,\rm{km/s}\,\rm{pc}^{2})$ seems to be applicable to a wide range of
galaxy formation models available in the literature. This paper is the
first to better quantify this by using the ASPECS data, the most
sensitive spectral survey currently available with ALMA, and properly
accounting for selection effects and survey area. The lack of \h2 in
$z>1$ model galaxies is possibly linked to a broader set of problems in galaxy formation and
evolution theory, for instance a lack of star-formation in galaxies
at these redshifts, and any solution should focus on all of these
simultaneously. We anticipate that additional surveys with ALMA and the JVLA,
focusing on larger and different areas in the sky and less \h2-rich galaxies will have the potential to further quantify the apparent lack of \h2 in galaxy
formation models, providing crucial additional constraints for our understanding of galaxy formation and evolution. These surveys should additionally address the conversion of an observed CO luminosity into an \h2 gas mass, while galaxy formation models should simultaneously focus on providing direct predictions for the CO luminosity of galaxies.

\acknowledgments
It is a pleasure to thank Ian Smail for comments on an earlier draft
of this paper. GP thanks Viraj Pandya, Adam Stevens, and Claudia Lagos
for useful discussions regarding the theoretical models discussed in
this work. {\gp The authors thank the referee for their constructive comments.} RSS and AY thank the Downsbrough family for their generous support, and gratefully acknowledge funding from the Simons Foundation. MV acknowledges support through an MIT RSC award, a Kavli Research Investment Fund, NASA ATP grant NNX17AG29G, and NSF grants AST-1814053 and AST-1814259. TDS acknowledges support from ALMA-CONICYT project 31130005 and FONDECYT project 1151239. JGL acknowledges partial support from ALMA-CONICYT project 31160033. DR acknowledges support from the National Science Foundation under grant number AST-1614213. This Paper makes use of the ALMA data ADS/JAO.ALMA\#2016.1.00324.L. ALMA is a partnership of ESO (representing its member states), NSF (USA) and NINS (Japan), together with NRC (Canada), NSC and ASIAA (Taiwan), and KASI (Republic of Korea), in cooperation with the Republic of Chile. The Joint ALMA Observatory is operated by ESO, AUI/NRAO and NAOJ. The National Radio Astronomy Observatory is a facility of the National Science Foundation operated under cooperative agreement by Associated Universities, Inc. Simulations for this work were performed on the Draco supercomputer at the Max Planck Computing and Data Facility, and on Rusty at the Center for Computational Astrophysics, Flatiron Institute.

\bibliographystyle{aasjournal.bst}
\bibliography{references}

\begin{figure*}
\center
\includegraphics[scale = 0.8]{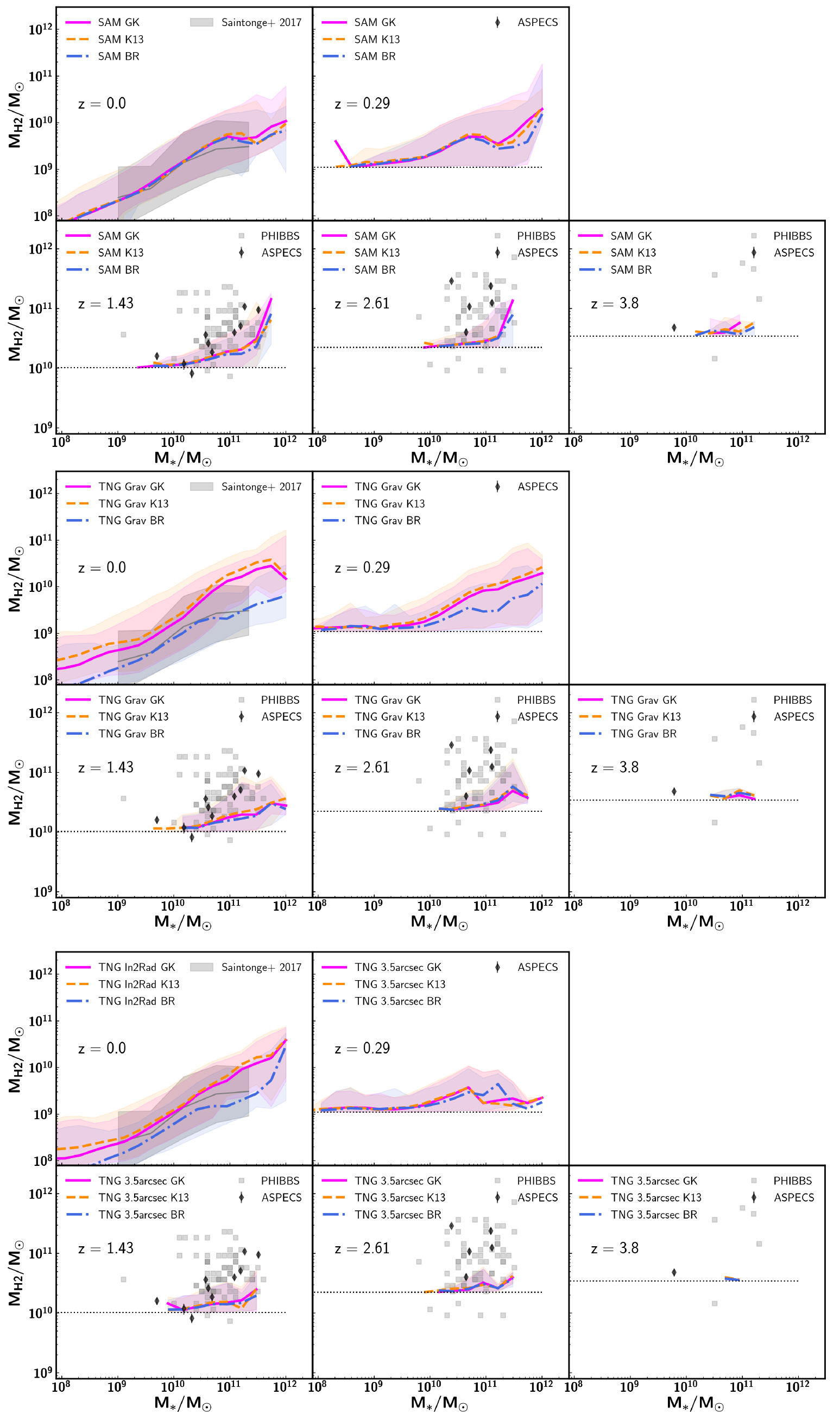}
\caption{
The \h2 mass of galaxies at different redshifts as a function of their
  stellar mass.  Model predictions are shown for the SC SAM (top two rows), IllustrisTNG adopting the `3.5arcsec' aperture (middle two rows, note that at $z=0$ this is replaced by then `In2Rad' aperture) and IllustrisTNG when adopting the `Grav' aperture (bottom two rows). For all models we show the GK
  (solid pink), the K13 (dashed orange), and the BR (dash-dotted blue)
  \h2 partitioning recipes. The selection function and data comparison is
  identical to Figure \ref{fig:scale_evol_SAMvsTNG}. The thick lines
  represent the median of the modeled galaxy population, whereas the
  shaded area represents the {\gp two-sigma scatter}. At $z>0$ the \h2 masses predicted by
  the different model variants are almost identical. The models are
  not able to reproduce all ASPECS and PHIBBS detections.
\label{fig:scale_evol_methods}}
\end{figure*}

\appendix
\section{Additional molecular hydrogen fraction recipes}
In Section \ref{sec:H2_recipes} we presented the relevant equations for the GK \h2 partitioning recipe. In this Section of the appendix we describe in detail the remaining two \h2 partitioning recipes adopted in this work.

\subsection{Blitz \& Rosolowsky 2006 (BR)}
\label{sec:BR}
The first recipe to calculate the molecular hydrogen fraction of the
cold gas in each cell is based on
the empirical pressure-based recipe presented by
\citet[BR]{Blitz2006}.  They found a power-law relation between the
disc mid-plane pressure and the ratio between molecular and atomic
hydrogen, i.e.,
\begin{equation}
R_{\mathrm{H}_2} = \left(\frac{\Sigma_{\mathrm{H}_2}}{\Sigma_{\mathrm{HI}}}\right) = \left(\frac{P_m}{P_0}\right)^{\alpha_{\rm BR}}
\label{eq:blitz2006}
\end{equation}
where $\Sigma_{\mathrm{H}_2}$ and $\Sigma_{\mathrm{HI}}$ are the \h2
and \hi surface densities, $P_0$ and $\alpha_{\rm BR}$ are free parameters that are
obtained from a fit to the observational data, and $P_m$ is the mid-plane
pressure acting on the galactic disc. 
We adopted $\log P_0/k_{\rm B} = 4.23$ cm$^{-3}$ K and $\alpha_{\rm BR}=0.8$
from \citet{Leroy2008}, where $k_{\rm B}$ is the Boltzmann constant. 

When calculating the \h2 fraction of cells in IllustrisTNG following the BR approach we replace the mid-plane pressure
$P_m$ by the thermal gas pressure of each cell, such that $P_m =
P_{th} = u\rho(\gamma -1)$.

We follow the approach described in PST14
and SPT15 to calculate the \h2 fraction of gas in the SC SAM. The mid-plane pressure is calculated as \citet{Elmegreen1989}:
\begin{equation}
P_{\rm m}(r) = \frac{\pi}{2}\,G\,\Sigma_{\mathrm{gas}}(r)\left(\Sigma_{\mathrm{gas}}(r) + f_{\sigma}(r)\Sigma_*(r)\right)
\label{eq:pressure}
\end{equation}
where G is the gravitational constant, $f_\sigma(r)$ is the
  ratio between $\sigma_{\mathrm{gas}}(r)$ and $\sigma_*(r)$, the gas
  and stellar vertical velocity dispersion, respectively. The stellar
  surface density profile $\Sigma_*(r)$ is modeled as an exponential
  with scale radius $r_{\mathrm{star}}$ and central density
  $\Sigma_{*, 0} \equiv m_*/(2 \pi r_*^2)$, where $m_*$ is the stellar mass of a galaxy.  Following \citet{Fu2012}, we adopt $f_{\sigma}(r) = 0.1 \sqrt{\Sigma_{*,0}/\Sigma_*}$.

\subsection{Krumholz 2013 (K13)}
\label{sec:K13}
The second recipe is based on the work presented in
\citet{Krumholz2013} and builds upon the works presented in
\citet{Krumholz2009,Krumholz2009_sf}. \citet{Krumholz2013} considers
an ISM that is composed by a warm neutral medium (WNM) and a cold
neutral medium (CNM) that are in pressure
equilibrium. \citet{Krumholz2013} finds that the equilibrium density of
this two-phase medium should be three times the minimum density and
writes
\begin{equation}
n_{\rm CNM,2p} = 3n_{\rm CNM,min} = 23 G_0\frac{4.1}{1 + 3.1
  (Z/Z_\odot)^{0.365}}\rm{cm}^{-3}.
\end{equation}
In the regime where the UV radiation field $G_0$ reaches zero, $n_{\rm
  CNM,2p}$ and the pressure also reach zero. This is an unphysical
scenario and to account for this \citet{Krumholz2013}  defines a minimum CNM density to maintain hydrostatic
balance, $n_{\rm CNM,hydro}$, based on the work by
\citet{Ostriker2010}.  This density depends on the thermal pressure
given as
\begin{equation}
 P_{\rm th} = 
\frac{\pi G\Sigma^2_{\rm HI}}{4\alpha} \,\times 
\biggl(1 + R_{\rm
  H2} +  2\sqrt{(1 + 2R_{\rm H2})^2 + 
  \frac{32\xi_d\alpha f_{\rm w}c^2_{\rm w}\rho_{\rm   sd}}{\pi G\Sigma^2_{\rm HI}}}\biggr),
\end{equation}
 where $R_{\rm H2} = \Sigma_{\rm H2}/\Sigma_{\rm HI}$,  $\alpha = 5$ describes how much of the mid-plane pressure support
is driven by turbulence, cosmic rays, and magnetic fields compared to
the thermal pressure, $\xi_d = 0.33$ is a geometrical factor, $c_{\rm w}
= 8\,\rm{km}\,\rm{s}^{-1}$ is the sound speed in the WNM, $f_{\rm w} = 0.5$ the ratio between the thermal velocity dispersion and
$c_{\rm w}$, and $\rho_{\rm sd} = 0.01\rm{M}_\odot\,\rm{pc}^{-3}$ the
stellar and dark matter density in the galactic disk.  $n_{\rm
  CNM,hydro}$ furthermore depends on the maximum temperature of the
CNM $T_{\rm CNM,max} = 243 \rm{K}$ \citep{Wolfire2003}, such that
\begin{equation}
n_{\rm  CNM,hydro} = \frac{P_{\rm th}}{1.1 \times k_{\rm B}T_{\rm CNM,max}}.
\end{equation}
The CNM density is then taken to be 
\begin{equation}
n_{\rm CNM} = \rm{max}(n_{\rm CNM,2p},n_{\rm CNM,hydro}).
\end{equation}
\citet{Krumholz2013} defines a dimensionless radiation field
\begin{equation}
\chi = 7.2 G_0\biggl(\frac{n_{\rm CNM}}{10\,\rm{cm}^{-3}}\biggr)^{-1}.
\end{equation}
The molecular hydrogen fraction is then given as
\begin{equation}
  f_{H_2}=\begin{cases}
     1 - 0.75s/(1 + 0.25s) & s <2\\
    0 & s\geq 2 
  \end{cases}
\end{equation}
where
\begin{equation}
s = \frac{\ln(1 + 0.6\chi + 0.01\chi^2)	}{0.6\tau_c}
\end{equation}
and
\begin{equation}
\tau_c = 0.066 f_c D_{\rm MW} \biggl(\frac{\Sigma_{\rm
    H}}{\rm{M}_\odot\,\rm{pc}^{-2}}\biggr).
\end{equation}
$f_c$ marks a clumping factor that accounts for the scale over which
the surface density is measured. The appropriate value for $f_c$
depends on the spatial scale over which the surface density is
measured and is suggested to be $f_c = 5$ on scales similar to the
resolution of IllustrisTNG. The same clumping factor is adopted for the SC SAM. The molecular hydrogen fraction $f_{H_2}$
depends on the ratio between the molecular and atomic surface density
$R_{\rm H2}$ and is solved iteratively.

\section{Predictions by different \h2 partitioning recipes}
\label{sec:partitioning_recipes}
We present the predictions for the \h2 mass of galaxies, \h2 mass function, and \h2 cosmic density adopting the three different \h2 partitioning recipes in this Appendix. Figure \ref{fig:scale_evol_methods} shows the \h2 mass as a function of stellar mass of galaxies after taking the ASPECS selection effects into account. These predictions are compared to the ASPECS results for the three different \h2 partitioning recipes adopted in this work, based on IllustrisTNG and the SC SAM, respectively. We find no difference in the predictions by the partitioning recipes for the SC SAM. When looking at IllustrisTNG we find that the BR partitioning recipe predicts \h2 masses that are systematically below the predictions by the other recipes at $z=0$ (0.1--0.2 dex). At higher redshifts the difference in the predictions by the three partitioning recipes is negligible.

\begin{figure*}
\center
\includegraphics[scale =0.8]{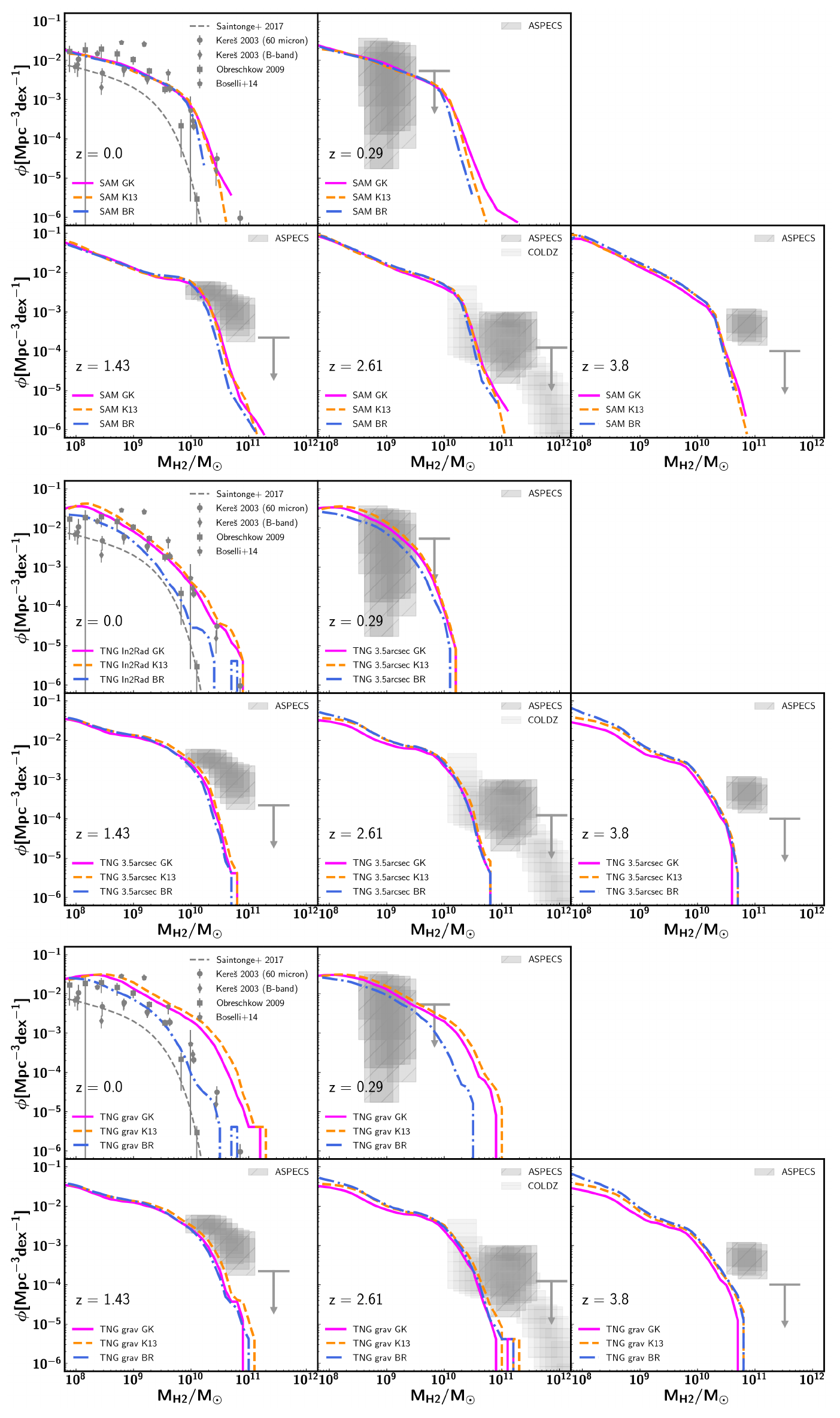}
\caption{The \h2 mass function of galaxies at $z=0$ and the redshifts probed by
  ASPECS. Model predictions are shown for the SC SAM (top two rows), IllustrisTNG adopting the `3.5arcsec' aperture (middle two rows, note that at $z=0$ this is replaced by then `In2Rad' aperture) and IllustrisTNG when adopting the `Grav' aperture (bottom two rows). For all models we show the GK
  (solid pink), the K13 (dashed orange), and the BR (dash-dotted blue)
  \h2 partitioning recipes. The data comparison is identical to Figure
  \ref{fig:massfunc_evol_SAMvsTNG}. \label{fig:massfunc_evol_methods}}
\end{figure*}

Figure \ref{fig:massfunc_evol_methods} shows the predictions from IllustrisTNG and the SC SAM for the \h2 mass function. When focusing on IllustrisTNG, the GK and K13 prescriptions result in almost identical \h2 mass functions at $z=0$. The
\citet{Obreschkow2009} observations are better reproduced when adopting
the BR \h2 partitioning recipe.  The BR prescription predicts 
number counts that are systematically below the predictions by the GK
and K13 prescriptions, the difference increasing to $\sim 0.5$ dex at
\h2 masses of $10^{10}\,\rm{M}_\odot$. At redshifts greater than zero the difference between the number densities predicted by the BR, GK, and K13 \h2 partitioning recipes decrease (at $z\sim 0.29$) or are minimal (at higher redshifts). Only in galaxies with \h2 masses less than $10^9\,\rm{M}_\odot$ at $z>3$ does the GK partitioning recipe predict number densities slightly less than the
other two recipes. This mass range is not covered by the ASPECS
survey. 

When we focus on the SC SAM we see that at $z=0$ the BR prescription predicts
slightly fewer galaxies with \h2 masses larger than
$10^{10}\,\rm{M}_\odot$ than the other prescriptions. The same is true
at $z=0.29$, whereas the predicted number densities are almost
identical for the \h2 mass functions at higher redshifts. 

\begin{figure*}
\includegraphics[width = \hsize]{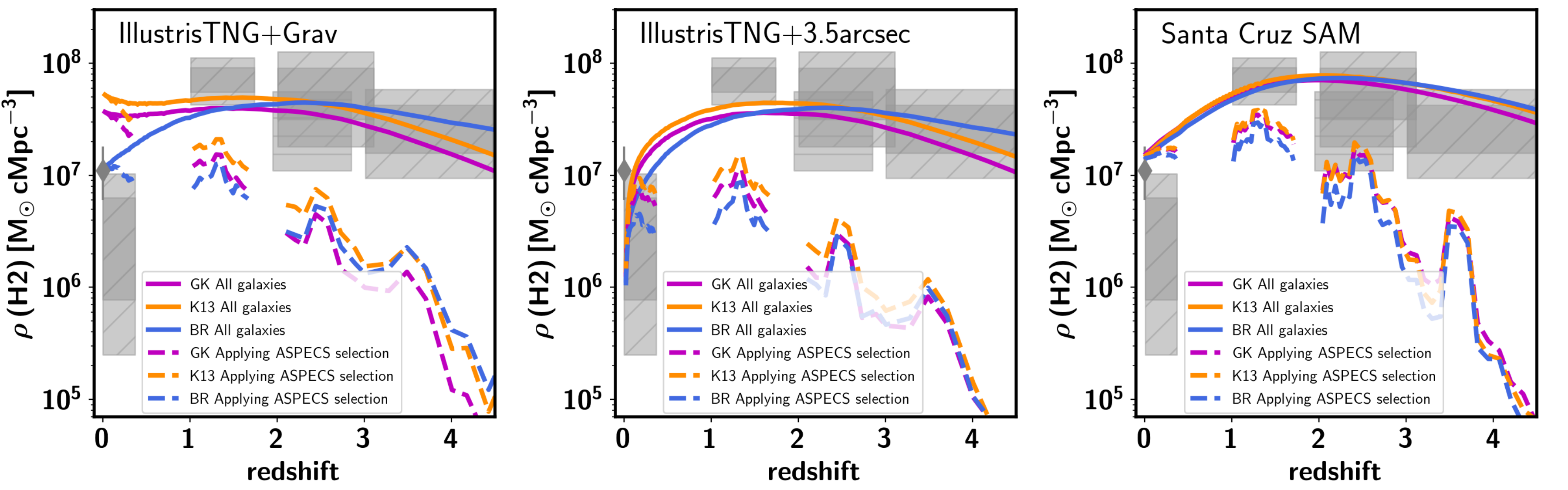}
\caption{Model predictions for the \h2 cosmic density as a function of
  redshift as predicted from IllustrisTNG when adopting the `Grav' aperture (left), IllustrisTNG when adopting the `3.5arcsec' aperture (middle), and the SC SAM (right). We show the 
   the GK (pink), K13 (orange), and BR (blue) \h2 partitioning
  recipes for all models. The solid lines correspond to the cosmic \h2 density in the entire box,
  ignoring any selection function. The dashed lines correspond to the
  cosmic \h2 density in the entire box, applying the ASPECS selection
  function. Model predictions are compared to the
  observations from ASPECS (dark (light) grey mark the
  one (two) sigma uncertainty), as well as observations at $z=0$ from \citet{Keres2003} and
  \citet{Obreschkow2009}.  \label{fig:rho_evol_SAMvsTNG_allmethods}}
\end{figure*}

We present the evolution of the \h2 cosmic density as predicted by
IllustrisTNG and the SC SAM in Figure \ref{fig:rho_evol_SAMvsTNG_allmethods} for all three \h2 partitioning recipes considered in this work (GK pink, K13 orange, BR blue). We find some differences in the predictions of the \h2 cosmic density by the different \h2 partitioning recipes for IllustrisTNG, mostly in the evolution of the \h2 cosmic density at redshifts $z<3$. The \h2 cosmic density gradually increases till $z=3$ for the BR partitioning recipe 
after which it decreases by a factor of 4 till $z=0$. The GK and K13
partitioning recipes predict a gradual increase in the cosmic density
till $z=2$, and a less pronounced decrease in the cosmic density till
$z=0$ of only a factor of $\sim 2$ for the `Grav' aperture. The predictions by the different partitioning recipes are similar when we account for the ASPECS sensitivity limits, typically within a
factor of 1.5 and at $z>1$. The \h2 cosmic density evolution predicted by the SC SAM is almost identical for the three partitioning recipes.  The cosmic densities predicted by the various partitioning recipes are also similar when accounting for the ASPECS selection function. This demonstrates that the systematic uncertainty between the different \h2 recipes is less than the typical uncertainty in the observations.

It is worthwhile to briefly focus on origin of the differences between the
different \h2 partitioning recipes. We demonstrated that the different
partitioning recipes yield almost identical predictions for the \h2
masses of galaxies as long as the underlying model is kept fixed. Only
at $z=0$ does the BR partitioning recipe coupled to IllustrisTNG
predict systematically lower \h2 masses. \citet{Diemer2018} also
demonstrated that the systematic uncertainty on average mass scaling
relations between different \h2
partitioning recipes coupled to IllustrisTNG is minimal (\citealt{Diemer2018} came to this conclusion exploring an even larger
sample of \h2 partitioning recipes). \citet{KrumholzGnedin2012} demonstrated that the GK and
\citet{Krumholz2009} recipes result in almost identical \h2
fractions. This is to first order driven by
the fact that both K13 and GK rely on the same set of input
parameters, primarily the surface density of neutral
hydrogen. Within the SC SAM, the BR recipe also primarily depends on
the surface density of neutral gas, which explains the negligible
differences in \h2 mass predictions by the different partitioning approaches. The BR recipe in the context
of the IllustrisTNG model is the only one that does not primarily depend on the neutral gas
surface density, but instead on the thermal
pressure. \citet{Diemer2018} argues that this implementation of the BR
partitioning recipe is incorrect, since the BR relation was calibrated
based on observations of the \hi and \h2 gas surface density in local
galaxies, rather than the thermal pressure as defined within
simulations. Despite this,
at $z>0$ the predictions between the different \h2 recipes for
IllustrisTNG are nearly identical.  It is furthermore curious that this approach yields
possibly best agreement with the observational data at $z=0$ (although
keep in mind that we did not properly mock the model predictions to
include observational selection and aperture effects).

\end{document}